\documentclass[superscriptaddress,amsmath,floatfix,aps,prc,twocolumn,10pt]{revtex4-2}
\pdfoutput=1
\def \dslash {\partial \! \! \! \slash}
\usepackage{preamblestyle} 
\usepackage{comment}

\newcommand{\neutrons}{n_{\pm}}
\newcommand{\protons}{p_{\pm}}
\newcommand{\neutron}{n_{+}}
\newcommand{\neutronMinus}{n_{-}}
\newcommand{\proton}{p_{+}}
\newcommand{\protonMinus}{p_{-}}
\newcommand{\nucleons}{N_{\pm}}
\newcommand{\nucleon}{N_{+}}
\newcommand{\nucleonMinus}{N_{-}}

\newcommand{\neutronDecayProton}{\neutron \rightarrow \proton + e^- + \bar{\nu}_e}
\newcommand{\neutronMinusDecayProton}{\neutronMinus \rightarrow \proton + e^- + \bar{\nu}_e}
\newcommand{\neutronDecayProtonMinus}{\neutron \rightarrow \protonMinus + e^- + \bar{\nu}_e}
\newcommand{\neutronMinusDecayProtonMinus}{\neutronMinus \rightarrow \protonMinus + e^- + \bar{\nu}_e}

\begin{document}

\title{Parity-doubled nucleons can rapidly cool neutron stars}
\author{Liam Brodie \orcidlink{0000-0001-7708-2073}}
\email{Corresponding author: b.liam@wustl.edu}
\affiliation{Department of Physics, Washington University in St.~Louis, St.~Louis, MO 63130, USA}
\affiliation{Department of Physics, Brookhaven National Laboratory, Upton, NY 11973}

\author{Robert D. Pisarski \orcidlink{0000-0002-7862-4759}}
\email{pisarski@bnl.gov}
\affiliation{Department of Physics, Brookhaven National Laboratory, Upton, NY 11973}

\date{October 23, 2025} 

\begin{abstract}
In confined hadronic matter, the spontaneous breaking and
restoration of chiral symmetry can be  
described by considering
nucleons, $\nucleon$(939), and excited states of opposite parity, $\nucleonMinus$(1535).  In a 
cold, dense hadronic phase
where chiral symmetry remains spontaneously broken, direct Urca decay processes involving the $\nucleonMinus$
are possible, e.g. $\nucleonMinus \rightarrow \nucleon + e^- + \bar{\nu}_e$.
We show that at low temperature
and moderate densities,
because the $\nucleonMinus$'s are much heavier than the $\nucleon$'s, such
cooling dominates over standard $\nucleon$ direct Urca
processes. This provides a strong astrophysical signature of the pattern of 
chiral symmetry restoration in neutron stars.
\end{abstract}

\maketitle

\noindent{\bf Introduction:}
\label{sec:intro}
The naive picture of phase transitions in quantum chromodynamics (QCD) is that there is just
a single transition from confined hadronic matter, in which chiral symmetry is
spontaneously broken, to a deconfined phase where chiral symmetry
is nearly restored.
Building upon numerical simulations of lattice QCD,
however, it is now understood that at zero quark
chemical potential and nonzero temperature, there is a wide range
of temperatures in which gluons and quarks are partially deconfined;
this can be described as a semi-quark-gluon-plasma \cite{Hidaka:2020vna}, or stringy liquid \cite{Glozman:2022lda}.
At low temperature and nonzero quark density, for a large number of colors
one can argue analytically that there is a quarkyonic phase,
which is confined but chirally symmetric until
high density~\cite{McLerran:2007qj}.

The basic model for baryons in
such a confined, chirally symmetric phase was
first given by Detar and Kunihiro \cite{Detar:1988kn,Jido:1998av,Jido:1999hd,Jido:2001nt,Zschiesche:2006zj,Dexheimer:2007tn,Dexheimer:2008cv,Gallas:2009qp,Parganlija:2010fz,Gallas:2011qp,Steinheimer:2011ea,Dexheimer:2012eu,Parganlija:2012fy,Holt:2014hma,Aarts:2017rrl,Mukherjee:2017jzi,Catillo:2018cyv,Marczenko:2018jui,Motornenko:2019arp,Marczenko:2019trv,Marczenko:2020jma,Minamikawa:2020jfj,Pisarski:2021aoz,Minamikawa:2021fln,Marczenko:2021uaj,Marczenko:2022hyt,Gao:2022klm,Minamikawa:2023eky,Kong:2023nue,Minamikawa:2023ypn,Koch:2023oez,Fraga:2023wtd,Eser:2023oii,Gao:2024mew,Gao:2024lzu,Giacosa:2024epf,Eser:2024xil,Yasui:2024dbx}. In such a model
the nucleons, $\nucleon (939)$, 
are considered with excited states of opposite parity, the
$\nucleonMinus (1535)$.  Doing so allows
the $\nucleon$'s and $\nucleonMinus$'s to have equal but nonzero
masses in a chirally symmetric phase.

Following previous work
\cite{Zschiesche:2006zj,Dexheimer:2008cv,Kong:2023nue,Fraga:2023wtd,Eser:2023oii,Eser:2024xil}
we introduce a subscript to denote the parity, so that $\neutron$ and $\proton$ are the usual
neutron and proton, with positive parity, 
while the excited states $\neutronMinus$ and $\protonMinus$ have negative parity.  Similarly,
$\nucleon$ denotes either $\neutron$ or $\proton$ and $\nucleonMinus$ denotes either $\neutronMinus$ or $\protonMinus$.

In this Letter we consider how parity doubled baryons affect
the cooling of neutron stars via neutrino emission. In a chirally symmetric phase, by definition the masses of the $\nucleon$ and $\nucleonMinus$'s are equal.  
In QCD, at densities several times that of nuclear saturation, chiral symmetry remains spontaneously broken. As a result, the $\nucleonMinus$ in-medium mass is still significantly greater than the $\nucleon$.
In many models of ordinary hadronic matter the direct Urca process is kinematically forbidden \cite{Steiner:2012rk,Alford:2023rgp},
and modified Urca processes, or contributions involving the width of the nucleon~\cite{Alford:2024xfb,Sedrakian:2024uma}, dominate. Our basic point is simple: with parity doubled nucleons, direct Urca processes from $\nucleonMinus \rightarrow \nucleon + e^- + \bar{\nu}_e$
open up at rather moderate densities, and that when they do,
they dominate direct and modified Urca processes of $\nucleon$'s
by {\it orders} of magnitude.

\noindent{\bf Microphysical inputs to emissivity:}
\label{app: pdm}
Our Lagrangian includes nucleons, mesons, and their interactions
with leptons through weak interactions:
\beq
\mathcal{L}=\mathcal{L}_N+\mathcal{L}_l+\mathcal{L}_M+\mathcal{L}_W \; .
\eeq
The nucleon Lagrangian is
\begin{align}
     \mathcal{L}_N = &\overline{\psi}_{1} \left( i \dslash 
    - g_{\omega} \omega \!\!\! \slash - g_{\rho} \vec{\rho} \!\!\! \slash \cdot \vec{\tau}
    - g_1 \left(\sigma + i \gamma_5 \vec{\pi} \cdot \vec{\tau} \right)
    \right) \psi_{1}
    \nonumber \\
    +  &\overline{\psi}_{2} \left( i \dslash 
    - g_{\omega} \omega \!\!\! \slash - g_{\rho} \vec{\rho} \!\!\! \slash \cdot \vec{\tau}
    - g_2 \left(\sigma - i \gamma_5 \vec{\pi} \cdot \vec{\tau}
    \right) \right) \psi_{2} 
    \nonumber \\
    & +  m_0 \left(
    \overline{\psi}_{2} \gamma_5 \psi_{1} - \overline{\psi}_{1} \gamma_5 \psi_{2} \right) \; .
\label{eq: nucleon_lagrangian}
\end{align}
We couple the nucleons to an $O(4)$ field $\phi = (\sigma, \vec{\pi})$,
as well as to the isosinglet vector meson, $\omega_\mu$, and the
isotriplet vector meson, $\vec{\rho}_\mu$.
The states $\psi_1$ and $\psi_2$ transform under $SU(2)_L\times SU(2)_R$ as 
\begin{align}
    \psi_{1L} &\rightarrow U_L\ \psi_{1L} && \psi_{1R} \rightarrow U_R\ \psi_{1R}\\
    \psi_{2L} &\rightarrow U_R\ \psi_{2L} && \psi_{2R} \rightarrow U_L\ \psi_{2R},
\end{align}
where $U_L$ and $U_R$ are elements of $SU(2)_L$ and $SU(2)_R$, respectively. By construction, the mass term, $m_0$,
is manifestly chirally symmetric.
When chiral symmetry is spontaneously broken by
an expectation value $\langle \sigma \rangle \neq 0$, the $\nucleon$ and $\nucleonMinus$ masses
are due both to $m_0$ and to their Yukawa couplings to the $\sigma$, with couplings constants
$g_1$ and $g_2$. Diagonalizing the mass matrix yields the mass eigenstates $\psi_{\nucleon}$ and $\psi_{\nucleonMinus}$ with masses
\begin{align}
    m_{\nucleons} = \pm \Big(\frac{g_1-g_2}{2}\Big) \sigma + \sqrt{ \Big(\frac{g_1+g_2}{2}\Big)^2 \sigma^2 +  m_0^2} \; .
    \label{eq: masses}
\end{align}

We assume that the axial $U(1)_A$ symmetry
is, as in vacuum, strongly broken quantum mechanically
by topologically nontrivial fluctuations.
If so, then the $\vec{a}_0$ and $\eta$ mesons, and their strange
counterparts, can be neglected
for the processes we consider.  It is
possible that the axial $U(1)_A$ symmetry is nearly restored
near the chiral phase transition ~\cite{Pisarski:2024esv,Giacosa:2024orp}, but we defer this analysis for now.

The meson part of the Lagrangian is 
\begin{align}
{\cal L}_M&=\frac{1}{2} (\partial_{\mu} \phi)^2 + 
\epsilon \sigma + \frac{\bar{\mu}^2}{2} \phi^2 - 
\frac{\lambda_4}{4}
\phi^4  + \frac{\lambda_6}{6}
\phi^6 \\
&+ \frac{1}{4} (F^\omega_{\mu \nu})^2 
+ \frac{1}{4} (\vec{F^\rho_{\mu \nu}})^2
+ \frac{m_\omega^2}{2} \omega_{\mu}^2  
+ \frac{m_\rho^2}{2} \vec{\rho}_{\mu}^{\, 2}
+ \lambda_{\omega \rho}\, \omega_\mu^2 \vec{\rho}_\mu^{\, 2} . 
\nonumber 
\end{align}
$F^\omega_{\mu \nu} = \partial_\mu \omega_\nu - \partial_\nu \omega_\mu$ and $\vec{F^\rho_{\mu \nu}} = \partial_\mu \vec{\rho}_\nu - \partial_\nu \vec{\rho}_\mu$
are the standard (Abelian) field strengths for the vector mesons.
We neglect a possible term $\sim \phi^2 \omega_\mu^2$ \cite{Pisarski:2021aoz}. This Lagrangian has been used in, for example, Ref.~\cite{Kong:2023nue}. The higher-order scalar meson self-interaction terms $\phi^4$ and $\phi^6$ are used to reproduce empirical properties of isospin-symmetric nuclear matter~\cite{Boguta:1977xi,Motohiro:2015taa}. The meson-meson interaction term $\omega_\mu^2\, \vec{\rho}_\mu^{\, 2}$ impacts the nuclear symmetry energy and allows for consistency with chiral effective field theory~\cite{Gao:2022klm,Malik:2024qjw}.

For the nucleon current which couples to the weak interactions, we take 
\begin{align}
    {\cal J}_\mu &= 
    \overline{\psi}_{\proton} (g_{V} - g_{A} \gamma_5) \gamma_\mu \psi_{\neutron} + \overline{\psi}_{\proton} (g_{V}^* - g_{A}^* \gamma_5) \gamma_\mu \psi_{\neutronMinus} \nonumber \\
    &+ \overline{\psi}_{\protonMinus} (g_{V}^* - g_{A}^* \gamma_5) \gamma_\mu \psi_{\neutron}
    + \overline{\psi}_{\protonMinus} (g_{V}^{**} - g_{A}^{**} \gamma_5) \gamma_\mu \psi_{\neutronMinus} \; .
    \label{nucleon_current}
\end{align}
We assume the vector couplings for the nucleons and their parity partners are $g_{V}=g_{V}^*=g_{V}^{**}=1$, 
and the axial vector coupling is $g_{A}=g_{A}^*=g_{A}^{**}=1.267$. 
The value of the axial couplings $g_A^*$ and $g_A^{**}$ are not strongly
constrained by either experiment or models.
Because the $\nucleonMinus$'s transform oppositely from
the $\nucleon$'s under chiral symmetry, we allow for negative values of $g_A^*$ and $g_A^{**}$. Therefore, the $\nucleonMinus$ can couple to a
right-handed current, instead of a left-handed current as the $\nucleon$'s do.
In fact this doesn't matter: the Fermi distribution of both fields
are symmetric for right and left-handed fields, so the decay amplitudes are the same.

The coupling constants we use come from Tables I, VI, and VII in Ref.~\cite{Kong:2023nue} for $m_0=600$ MeV and $L=40$ MeV: $g_1=8.48,\ g_2=14.93,\ \epsilon=1.81104\times10^6\ \text{MeV}^3,\ \lambda_4=40.39,\ \lambda_6=0.00184475\ \text{MeV}^{-2},\ g_{\omega}=9.13,\ g_{\rho}=10.99,\ \lambda_{\omega \rho}=862.815,\ \overline{\mu}=436.828\ \text{MeV},\ m_{\omega}=783\ \text{MeV},\ m_{\rho}=776\ \text{MeV}$.  This set of parameters produces an equation of state that satisfies three criteria:
first, it is consistent with properties of isospin-symmetric nuclear matter near nuclear saturation density; 
second, it is consistent with chiral effective field theory predictions about the binding energy of neutron matter from Ref.~\cite{Drischler:2020yad}; third, it predicts a maximum mass of $M = 2.19$ M$_{\odot}$ and $R(1.4\ \text{M}_{\odot})=12.9$ km,
in accord with present observations~\cite{Miller:2019cac,Riley:2019yda,Miller:2021qha,Riley:2021pdl,Choudhury:2024xbk,Salmi:2022cgy,Salmi:2024aum,Dittmann:2024mbo,Fonseca:2021wxt,Vinciguerra:2023qxq}. We work in the mean-field approximation, details can be found in Ch.~3 of Ref.~\cite{Schmitt:2010pn} and in Ref.~\cite{Kong:2023nue}.

\noindent{\bf Flavor-changing processes:}
We focus on the cooling of neutron stars, assuming that the
temperature is sufficiently low such that the neutrino mean free path is at
least as large as the radius of the star. Through the weak interactions, the coupling of the charged currents 
in Eq.~\ref{nucleon_current} generate eight flavor-changing processes:
\begin{align}
    &\neutron \rightarrow \proton + e^{-} + \bar{\nu}_e &&\proton + e^{-} \rightarrow \neutron + \nu_{e} \label{eq: nd}\\
    &\neutronMinus \rightarrow \proton + e^{-} + \bar{\nu}_e &&\proton + e^{-} \rightarrow \neutronMinus + \nu_{e} \label{eq: n^*d}\\
    &\neutron \rightarrow \protonMinus + e^{-} + \bar{\nu}_e &&\protonMinus + e^{-} \rightarrow \neutron + \nu_{e} \label{eq: nd(p^*)}\\
    &\neutronMinus \rightarrow \protonMinus + e^{-} + \bar{\nu}_e  &&\protonMinus + e^{-} \rightarrow \neutronMinus + \nu_{e}. \label{eq: n^*d(p^*)}
\end{align}
We consider matter in chemical equilibrium ($\mu_{\neutron}=\mu_{\proton}+\mu_{e^-},\; \mu_{\neutron}=\mu_{\neutronMinus},\; \mu_{\proton}=\mu_{\protonMinus}$),
so each process of neutron decay and its corresponding process of electron capture are in equilibrium,
leaving four independent direct Urca processes. If we considered higher temperatures, such as for the merger of
two neutron stars, we would also have to consider the eight inverse processes of Eq.~\ref{eq: nd} -- Eq.~\ref{eq: n^*d(p^*)}.

The neutrino emission rate from a star has dimensions of energy
per spacetime volume, and is often called the emissivity.
It can be computed using Eq.~(120) of Ref.~\cite{Yakovlev:2000jp} and Eq.~(7) of Ref.~\cite{Lattimer:1991ib}. 
At low temperatures, $T \lesssim 1$~MeV, the emissivity is dominated by particles near their Fermi surfaces. The sum of emissivities for each pair of neutron decay and electron capture direct Urca (dU) processes is
\beq
Q^{\text{dU}}=\frac{457\ \pi}{10080} G_F^2 \cos^2{\theta_C}(1+3\,g_{A}^2)m_{\neutrons} m_{\protons} m_e T^6 \Theta^{\text{dU}}.
\label{eq: du_emissivity}
\eeq
Here $G_F = 1.16637 \times 10^{-11}$ MeV$^{-2}$, 
the Cabibbo angle $\theta_C = 13.02^{\circ}$, and $g_{A}= 1.267$.
Because of the uncertainty in parity doubled models for the value
of $g_A^*$ and $g_A^{**}$, we set them equal to $g_A$. 
The in-medium particle masses are denoted $m_{\pm}$, $T$ is the temperature, and $\Theta^{\text{dU}}=1$ if the direct Urca process is kinematically 
allowed, and $=0$ if not. At low temperatures, the direct Urca process is allowed if the $\protons$ and $e^{-}$ Fermi momenta are greater than the $\neutrons$ Fermi momentum, $k^{F}_{\protons}+k^{F}_{e^-}\geq k^{F}_{\neutrons}$. When direct Urca processes
are not allowed, modified Urca processes are relevant,
\begin{align}
\label{eq: mu_reactions}
    \neutron + \nucleon &\rightarrow \proton + \nucleon + e^{-} + \bar{\nu}_e \nonumber \\
    \proton + \nucleon + e^{-} &\rightarrow \neutron + \nucleon + \nu_e \; .
\end{align}
We only consider these modified Urca (mU) processes because, in this parity doublet model, $\neutronDecayProton$, Eq.~\ref{eq: nd}, is kinematically forbidden. When $\nucleon=\neutron$ in Eq.~\ref{eq: mu_reactions}, the modified Urca contribution to the emissivity 
is given by Eq.~(140) in Ref.~\cite{Yakovlev:2000jp} and Eq.~(65c) in Ref.~\cite{Friman:1979ecl}:
\beq
Q^{\text{mU,}\neutron} = A\, G_F^2 \cos^2{\theta_C} g_A^2 
\frac{m_{\neutron}^3 m_{\proton} k^{F}_{\proton}}{m_{\pi}^4} T^8 \; ,
\eeq
where $m_{\pi} = 139$~MeV is
the pion mass and the constant $A = 0.04656$ \cite{Friman:1979ecl}.

When $\nucleon = \proton$ in Eq.~\ref{eq: mu_reactions}, the modified Urca contribution to the emissivity follows from Eq.~142 of Ref.~\cite{Yakovlev:2000jp},
and is
\beq
Q^{\text{mU,}\proton} \approx Q^{\text{mU,}\neutron} \frac{m_{\proton}^2}{m_{\neutron}^2} \frac{(k^{F}_{e^-}+3k^{F}_{\proton}-k^{F}_{\neutron})^2}{8 k^{F}_{e^-}\ k^{F}_{\proton}} \, \Theta^{\text{mU,}\proton},
\eeq
where $\Theta^{\text{mU,}\proton}$ is 1 if $k^{F}_{e^-}+3k^{F}_{\proton}>k^{F}_{\neutron}$ and 0 otherwise. 

To make an estimate about the cooling capabilities of flavor-changing processes involving the nucleon's parity partner we consider the heat lost due to neutrino emission
\beq
Q(T) = -c_V(T)\ \frac{dT}{dt},
\eeq
where $c_V$ is the specific heat at constant volume ($c_V(T)=T\ ds/dT|_{T=0}$ with entropy density $s$) and $t$ is the time. To find $T(t)$, we first integrate this expression with respect to an initial time $t_0$ and temperature $T_0$, as in Ch.~5.3 of Ref.~\cite{Schmitt:2010pn}, to find $t(T)$
\beq
\int_{t_0}^{t}dt^{'} = -(ds/dT|_{T=0}) \int_{T_0}^{T} dT^{'} \frac{T^{'}}{\widetilde{Q}^{\text{dU}}(T^{'})^6 + \widetilde{Q}^{\text{mU}}(T^{'})^8},
\eeq
where $\widetilde{Q}$ is the coefficient in $Q = \widetilde{Q}\, T^n$. We can then invert this expression to get $T(t)$. After the birth of a neutron star, there is a period of thermal relaxation ($t \lesssim 10-50$ years), then neutrino emission becomes the dominant cooling mechanism over photon emission ($t\lesssim 10^5$ years) \cite{Yakovlev:2003qy}. To capture the density dependence of the neutrino emissivity throughout the star, we integrate over the volume of the star from the center to the crust assuming the temperature is constant. The radial dependence of $Q$ and $c_V$ are found by computing the star's radial density profile using the Tolman-Oppenheimer-Volkoff (TOV) equation from general relativity. For the purpose of computing stellar properties using the TOV equation, we attach the GPPVA(TM1e) crust equation of state \cite{Grill:2014aea,tm1e_crust} at baryon chemical potential $\mu_B = 952\ \text{MeV}$, which corresponds to a baryon number density of $n_B=0.07\, \text{fm}^{-3}$ in the core, see Ref.~\cite{Alford:2022bpp} for details about the attachment procedure.

\begin{figure}
\includegraphics[width=0.48\textwidth]{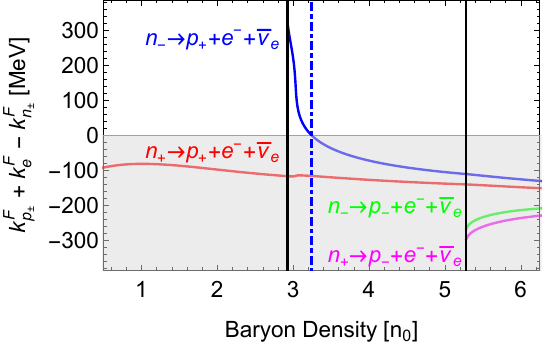}
\caption{The deficit of momentum forbidding the direct Urca process as a function of baryon number density. A Fermi sea of $\neutronMinus$'s appears at $n_B=2.94\ \nsat$;
for $\protonMinus$'s, at $n_B=5.31\ \nsat$. The dot-dashed line indicates where the $\neutronMinusDecayProton$ process terminates at $n_B = 3.25\ \nsat$. The grey region indicates where direct Urca is kinematically forbidden.
}
\label{fig: momentum_deficit}
\end{figure}

\noindent{\bf Results:}
\label{sec:results}
In Fig.~\ref{fig: momentum_deficit} we show the difference in particle momenta such that negative values mean there is a deficit of momentum, forbidding the direct Urca process. The direct Urca process is kinematically allowed at low temperatures ($T\lesssim 1\ \text{MeV}$) if the sum of $\protons$ and $e^-$ Fermi momenta is greater than the $\neutrons$ momentum, $k^{F}_{\protons}+ k^{F}_{e^-} \geq k^{F}_{\neutrons}$. Note that when the number of $e^-$ and $\proton$ are not equal, such as when there is a nonzero population of $\protonMinus$, two other kinematic conditions must be satisfied: $k^{F}_{e^-} + k^{F}_{\neutrons} \geq k^{F}_{\protons}$ and $k^{F}_{\protons} +  k^{F}_{\neutrons} \geq k^{F}_{e^-}$. There is no neutrino Fermi surface at low temperatures because the mean free path of the neutrinos is longer than the size of the neutron star; the average neutrino momentum is similar to the temperature, and can therefore be neglected. 

For ordinary hadronic matter, involving $\neutron$'s, $\proton$'s, and $e^-$'s,
there is a threshold for direct Urca processes.  This is because, at low
temperature, all momenta that participate in Urca processes are near the Fermi surface, so
for $\neutronDecayProton$
to proceed, the momenta must satisfy the kinetic constraint 
$k^{F}_{\proton}+ k^{F}_{e^-} \geq k^{F}_{\neutron}$.  This is fulfilled for all densities above the direct Urca threshold, where the fraction of protons to the total number of baryons is greater than $\approx 11 \%$ \cite{Lattimer:1991ib}. In parity doublet models, there is a threshold when
$\neutronMinus$'s first form a Fermi sea, and the process $\neutronMinusDecayProton$ opens up.  
This terminates when the kinematic 
constraint for $\neutronMinus$'s to decay,
$k^{F}_{\proton}+ k^{F}_{e^-} \geq k^{F}_{\neutronMinus}$, cannot be fulfilled, and 
$\neutronMinus$ decay turns off.  
This is indicated by the dot-dashed line in Fig.~\ref{fig: momentum_deficit}.
This termination is expected: in the chirally symmetric regime
$\nucleon$'s and $\nucleonMinus$'s are degenerate, and standard Urca
processes turn off.  Unremarkably, this happens before one fully reaches
the chirally symmetric regime, as the $\nucleon$'s and $\nucleonMinus$'s 
become closer in mass.

The direct Urca processes $\neutronDecayProton$, Eq.~\ref{eq: nd} and $\neutronDecayProtonMinus$, Eq.~\ref{eq: nd(p^*)}, are never allowed because there are too many $\neutron$'s. We do not mention the corresponding electron capture processes because their emissivity is equal to the neutron decay emissivity in chemical equilibrium. A Fermi sea of 
$\neutronMinus$'s forms when $n_B = 2.94\ \nsat$, where $n_B$ is the number of baryons per unit volume and $\nsat =0.16\ \text{fm}^{-3}$ is nuclear saturation density,
and is where the process $\neutronMinus \rightarrow \proton + e^- + \bar{\nu}_e$, Eq.~\ref{eq: n^*d}, 
is first allowed.
This process becomes kinematically forbidden when $n_B =3.25\ \nsat$, 
as the $\neutronMinus$ mass decreases, and the density of $\neutronMinus$'s
increases. This process is similar to the strangeness-changing process $\Lambda_+ \rightarrow \proton + e^- + \bar{\nu}_e$;
only a tiny fraction of $\Lambda_+$ are needed for the process to be kinematically allowed~\cite{Prakash:1992zng}, but there is no termination density.
The onset density for $\protonMinus$'s is $n_B = 5.31\ \nsat$. The process $\neutronMinusDecayProtonMinus$, Eq.~\ref{eq: n^*d(p^*)}, is never allowed because there are too many $\neutronMinus$'s. 

The precise values for these thresholds are
clearly sensitive to the details of our model such as the values of the coupling constants and chirally invariant mass, which degrees of freedom
are included, and the mean-field approximation.

\begin{figure}
\includegraphics[width=0.48\textwidth]{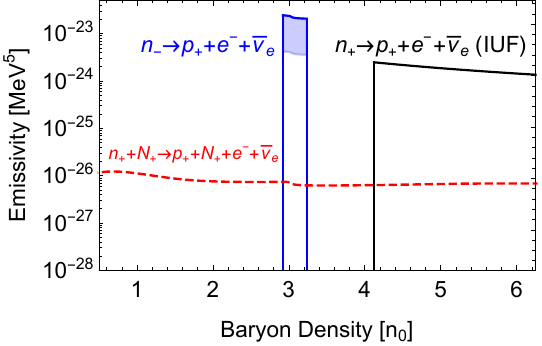}
\caption{The rate of neutrino emission at $T=100$ keV as a function of baryon number density. Direct Urca emissivities are in solid colors and modified Urca emissivities are dashed. 
We vary $g_A^*$ from $-1.267$ to $+1.267$, and indicate the effect by the blue shaded region. In our model, $\neutronMinusDecayProton$ is allowed between the onset of
$\neutronMinus$'s at $n_B = 2.94\ \nsat$ and their termination at
$n_B = 3.25\ \nsat$.
Since $\neutronDecayProton$ is not allowed in our model, for comparison
we include the results from a different model, the
IUF model of Ref.~\cite{Fattoyev:2010mx}.
}
\label{fig: emissivity}
\end{figure}

In Fig.~\ref{fig: emissivity}, we show the neutrino emissivity as a function of baryon number density. We compare the emissivities from the $\neutronMinusDecayProton$ direct Urca process,
the modified Urca process of Eq.~\ref{eq: mu_reactions},
and the $\neutronDecayProton$ direct Urca from the Indiana University and Florida State University (IUF) relativistic mean-field theory \cite{Fattoyev:2010mx}. Over the region where $\neutronMinusDecayProton$ is allowed in Fig.~\ref{fig: momentum_deficit}, its emissivity is orders of magnitude larger than the modified Urca process and 
the IUF $\neutronDecayProton$ direct Urca process
(which has a direct Urca threshold at $n_B = 4.13\ \nsat$). This is primarily due to the elementary fact that the $\neutronMinus$
in-medium mass is greater than the $\neutron$, and that the direct Urca emissivity, Eq.~\ref{eq: du_emissivity}, is directly dependent upon this quantity. The blue-shaded region shows the effect of varying $g_A^*$ from $-g_A$ to $+g_A$. We only consider the modified Urca process of Eq.~\ref{eq: mu_reactions} because when $\neutronMinusDecayProton$ is allowed, modified Urca contributions from $\nucleonMinus$ processes are negligible. 

\begin{figure}
\includegraphics[width=0.48\textwidth]{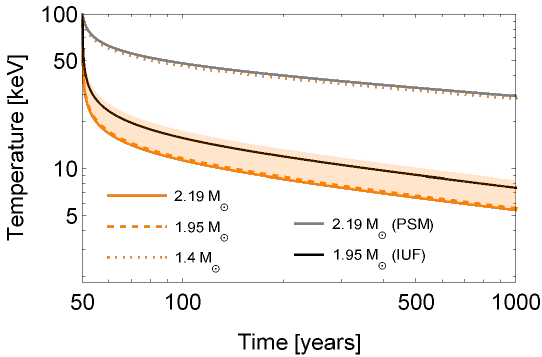}
\caption{
The core temperature of various mass neutron stars over time since birth. The $\neutronMinus$ onset density is only reached in the $1.95\ \text{M}_{\odot}$ and $2.19\ \text{M}_{\odot}$ stars (dashed and solid orange curves) and therefore those stars cool faster due to the \mbox{$\neutronMinusDecayProton$} process. For the $2.19\ \text{M}_{\odot}$ star we vary $g_A^*$; the effect is shown with the orange shaded region. The parity singlet model (PSM) and IUF model only include $\nucleon$ and cool primarily by modified and direct Urca processes, respectively. 
}
\label{fig: temp_of_time}
\end{figure}

In Fig.~\ref{fig: temp_of_time}, we show how the internal temperature of stars with different masses vary over time. The initial temperature for all the stars is 100 keV. We assume that the end of the thermal relaxation period is 50 years \cite{Yakovlev:2003qy}. We compare two $M=2.19\ \text{M}_{\odot}$ stars -- one with parity doubled baryons and one with only $\nucleon$, denoted PSM for the parity singlet model, which comes from the same Lagrangian, Eq.~\ref{eq: nucleon_lagrangian}, with $\mu_{\nucleonMinus}=0$. Due to \mbox{$\neutronMinusDecayProton$}, the star with parity doubled baryons cools {\it much} quicker, with characteristic cooling timescales of 159 {\it days} compared with 298 {\it years}. The orange shaded region shows the effect of varying $g_A^*$ from $-g_A$ to $+g_A$ for the $2.19\ \text{M}_{\odot}$ star with $\nucleonMinus$. Note that the mass range corresponding to the range of central densities where \mbox{$\neutronMinusDecayProton$} is allowed ($2.94\,\nsat - 3.25\, \nsat$) is $1.83\,\text{M}_{\odot}-1.97\,\text{M}_{\odot}$. We note that the values of the central density window and quoted mass range are specific to the model we use. 
In different models, such as those within Ref.~\cite{Kong:2023nue},
these values will change. For the models developed in Ref.~\cite{Steinheimer:2025hsr}, the phase space for the $\neutronMinusDecayProton$ process opens up \cite{PDMCoolingProj}. 
We therefore suggest that the opening of Urca processes in a finite density window from parity doubled nucleons could well be a robust phenomenon, even though the precise central density and mass ranges depend on the chosen model.

We also compare the parity doublet model with the IUF relativistic mean-field theory, which only includes $\nucleon$, for a $M=1.95\ \text{M}_{\odot}$ star (the maximum mass predicted by the IUF model). Direct Urca is only allowed between $2.93\leq r \leq 4.4\ \text{km}$ in the parity doubled star compared to $0\leq r \leq 5.1\ \text{km}$ in the IUF $\nucleon$ star, but the parity doubled star cools quicker, with characteristic cooling timescales of 179 days for parity doubled baryons and 236 days for IUF $\nucleon$'s. This is due to the enhanced emissivity for $\neutronMinusDecayProton$ shown in Fig.~\ref{fig: emissivity}. The $M=1.4\ \text{M}_{\odot}$ star with parity doubled baryons has a central density less than the $\nucleonMinus$ onset density, so this star cools slowly by the modified Urca process of Eq.~\ref{eq: mu_reactions}.

The data underlying Figs.~\ref{fig: momentum_deficit}, \ref{fig: emissivity}, and \ref{fig: temp_of_time} are available in Ref.~\cite{BrodiePisarski2025Data}.

\noindent{\bf Conclusions:}
\label{sec:conclusions}
Our basic assumption in this Letter is that for densities relevant to neutron stars, hadronic matter is described by confined baryons, in a phase in which chiral symmetry
remains spontaneously broken.
In such a phase, because of the mass splitting between
the $\nucleonMinus$ and
the $\nucleon$, neutrino emission from direct Urca processes involving $\nucleonMinus$'s dominates over direct Urca only involving the $\nucleon$. 

We note that neutrino emission in parity doubled models was considered previously \cite{Dexheimer:2012eu,Mukherjee:2017jzi,Marczenko:2018jui}.
Ref.~\cite{Dexheimer:2012eu} considered the effect the population of $\nucleonMinus$ has on
the direct Urca threshold and emissivity for processes only involving the $\nucleon$. Ref.~\cite{Mukherjee:2017jzi} mentioned the lack of a direct Urca threshold for processes involving the $\nucleon$.
Ref.~\cite{Marczenko:2018jui} noted when direct Urca processes
involving the $\nucleonMinus$ are kinematically allowed in the chirally symmetric phase. 
Our analysis is the first to compute the direct Urca neutrino emissivity for processes involving the $\nucleonMinus$.

We stress that the (approximate) restoration of chiral symmetry is an inevitable
feature of QCD at high temperature and/or baryon density.
For example, 
numerical simulations of lattice QCD find that the $\nucleon$ and $\nucleonMinus$ masses become
degenerate at zero chemical potential near the critical temperature~\cite{Aarts:2017rrl}. 
Thus at low temperature and nonzero baryon density, 
even in a phase where the chiral symmetry remains spontaneously
broken, including the $\nucleonMinus$'s in a parity
doubled model is most natural.

Flavor-changing weak interaction processes involving parity doubled baryons 
can also impact the dynamics of matter in a neutron star merger. If the flavor relaxation timescale is similar to the frequency of density oscillations in a merger, bulk viscous effects arise which can damp density oscillations and therefore impact the postmerger gravitational-wave signal \cite{Alford:2017rxf,Most:2022yhe}. The enhanced neutrino emissivity in the finite density interval where \mbox{$\neutronMinusDecayProton$} is allowed can affect other astrophysical signals such as the kilonova.

The mechanism proposed here may be distinguished from the standard direct Urca cooling if we had a sufficient sample of young neutron stars. Since we do not, one other method is to look at stars in thermal equilibrium, e.g., SAX J1808.4-3658 and 1H 1905+000 \cite{Heinke:2006ie,Jonker:2006td} to see if the proposed \mbox{$\neutronMinusDecayProton$} direct Urca process can be distinguished from the standard direct Urca process \mbox{$\neutronDecayProton$} using the upper bound on the thermal emission from these stars. We plan to pursue this in a separate work.

Our present model has obvious limitations.  First,
we consider two light flavors instead of the physically realistic case of $2+1$ flavors. It is essential to know
when a Fermi sea of strange baryons forms and affects the
equation of state.
Similarly, in vacuum the $\nucleonMinus(1535)$ decays
to $\nucleon \eta$ \cite{ParticleDataGroup:2024cfk},
so the $\nucleonMinus$ clearly couples to strange quarks. 

Second, instead of using $g_A$ from the model, we kept it as a free
parameter.  However, in parity doubled models at tree level 
the axial coupling constant $g_A$ is less than unity.
This can be ameliorated by coupling to mesons with spin-$1$
with a more involved analysis
\cite{Gallas:2009qp,Parganlija:2010fz,Parganlija:2012fy}.

Lastly, we have neglected pairing.  This
can be the usual nucleon gaps,  
$\langle \nucleon \nucleon \rangle$
\cite{Page:2009fu,Beloin:2016zop,Sedrakian:2018ydt}, as well as 
$\langle \nucleonMinus \nucleonMinus \rangle$.
(Cross pairing, 
$\langle \nucleon \nucleonMinus \rangle$,
is suppressed because the Fermi momenta for $\nucleon$ and
$\nucleonMinus$ differ in the chirally broken phase.)
$\langle \nucleon \nucleon \rangle$ pairing is familiar; $\langle \nucleonMinus \nucleonMinus \rangle$ pairing
presumably occurs, but as the Fermi momentum of the $\nucleonMinus$
is much smaller than the $\nucleon$ for the density range where the $\nucleonMinus$ neutrino emissivity is nonzero, presumably so are $\nucleonMinus$ gaps (see Eq.~(11) in Ref.~\cite{Sedrakian:2018ydt} and Eq.~(18) in Ref.~\cite{Page:2013hxa}). Assuming that the $\neutronMinus$ gap is small compared to the $\neutron$ gap
implies that neutrino emission due to $\neutronMinusDecayProton$ is enhanced even more compared to standard neutron decay $\neutronDecayProton$, which is
suppressed by $\neutron$ pairing.

The fact that direct Urca processes can open up for $\nucleonMinus$ is,
we suggest, an inescapable aspect of neutron stars. The value of the $\nucleonMinus$ onset density and the range of densities where $\nucleonMinus$ neutrino emissivity is nonzero may be model dependent, but the phenomenon of enhanced neutrino emissivity due to $\neutronMinusDecayProton$ is general.
This might help to explain the examples
of neutron stars which appear to cool more
rapidly than by standard mechanisms \cite{Brown:2017gxd,Marino:2024gpm}.

\acknowledgments
L.B. and R.D.P. thank Mark G.~Alford, Veronica Dexheimer, Alexander Haber, Teiji Kunihiro, Sanjay Reddy, Chihiro Sasaki, Lorenz von Smekal, and Wolfram Weise for useful discussions. R.D.P. thanks the long-term workshop, HHIQCD2024, at the Yukawa Institute for Theoretical Physics (YITP-T-24-02), for
their hospitality and many useful talks.  R.D.P. is supported by the
U.S. Department of Energy under contract DE-SC0012704, and thanks
the Alexander von Humboldt Foundation for their support.  
L.B. is partly supported by the U.S. Department of Energy, Office of Science, Office of Nuclear Physics, under Award
No. DE-FG02-05ER41375 and partly supported by the U.S. Department of Energy, Office of Science, Office of Workforce Development for Teachers and Scientists, Office of Science Graduate Student Research (SCGSR) program. The SCGSR program is administered by the Oak Ridge Institute for Science and Education for the DOE under contract number DE‐SC0014664. 

%
%


\bibliography{reflist}

\begin{thebibliography}{85}%
\makeatletter
\providecommand \@ifxundefined [1]{%
 \@ifx{#1\undefined}
}%
\providecommand \@ifnum [1]{%
 \ifnum #1\expandafter \@firstoftwo
 \else \expandafter \@secondoftwo
 \fi
}%
\providecommand \@ifx [1]{%
 \ifx #1\expandafter \@firstoftwo
 \else \expandafter \@secondoftwo
 \fi
}%
\providecommand \natexlab [1]{#1}%
\providecommand \enquote  [1]{``#1''}%
\providecommand \bibnamefont  [1]{#1}%
\providecommand \bibfnamefont [1]{#1}%
\providecommand \citenamefont [1]{#1}%
\providecommand \href@noop [0]{\@secondoftwo}%
\providecommand \href [0]{\begingroup \@sanitize@url \@href}%
\providecommand \@href[1]{\@@startlink{#1}\@@href}%
\providecommand \@@href[1]{\endgroup#1\@@endlink}%
\providecommand \@sanitize@url [0]{\catcode `\\12\catcode `\$12\catcode `\&12\catcode `\#12\catcode `\^12\catcode `\_12\catcode `\%12\relax}%
\providecommand \@@startlink[1]{}%
\providecommand \@@endlink[0]{}%
\providecommand \url  [0]{\begingroup\@sanitize@url \@url }%
\providecommand \@url [1]{\endgroup\@href {#1}{\urlprefix }}%
\providecommand \urlprefix  [0]{URL }%
\providecommand \Eprint [0]{\href }%
\providecommand \doibase [0]{https://doi.org/}%
\providecommand \selectlanguage [0]{\@gobble}%
\providecommand \bibinfo  [0]{\@secondoftwo}%
\providecommand \bibfield  [0]{\@secondoftwo}%
\providecommand \translation [1]{[#1]}%
\providecommand \BibitemOpen [0]{}%
\providecommand \bibitemStop [0]{}%
\providecommand \bibitemNoStop [0]{.\EOS\space}%
\providecommand \EOS [0]{\spacefactor3000\relax}%
\providecommand \BibitemShut  [1]{\csname bibitem#1\endcsname}%
\let\auto@bib@innerbib\@empty
\bibitem [{\citenamefont {Hidaka}\ and\ \citenamefont {Pisarski}(2021)}]{Hidaka:2020vna}%
  \BibitemOpen
  \bibfield  {author} {\bibinfo {author} {\bibfnamefont {Y.}~\bibnamefont {Hidaka}}\ and\ \bibinfo {author} {\bibfnamefont {R.~D.}\ \bibnamefont {Pisarski}},\ }\bibfield  {title} {\bibinfo {title} {{Effective models of a semi-quark-gluon plasma}},\ }\href {https://doi.org/10.1103/PhysRevD.104.074036} {\bibfield  {journal} {\bibinfo  {journal} {Phys. Rev. D}\ }\textbf {\bibinfo {volume} {104}},\ \bibinfo {pages} {074036} (\bibinfo {year} {2021})},\ \Eprint {https://arxiv.org/abs/2009.03903} {arXiv:2009.03903 [hep-ph]} \BibitemShut {NoStop}%
\bibitem [{\citenamefont {Glozman}\ \emph {et~al.}(2022)\citenamefont {Glozman}, \citenamefont {Philipsen},\ and\ \citenamefont {Pisarski}}]{Glozman:2022lda}%
  \BibitemOpen
  \bibfield  {author} {\bibinfo {author} {\bibfnamefont {L.~Y.}\ \bibnamefont {Glozman}}, \bibinfo {author} {\bibfnamefont {O.}~\bibnamefont {Philipsen}},\ and\ \bibinfo {author} {\bibfnamefont {R.~D.}\ \bibnamefont {Pisarski}},\ }\bibfield  {title} {\bibinfo {title} {{Chiral spin symmetry and the QCD phase diagram}},\ }\href {https://doi.org/10.1140/epja/s10050-022-00895-4} {\bibfield  {journal} {\bibinfo  {journal} {Eur. Phys. J. A}\ }\textbf {\bibinfo {volume} {58}},\ \bibinfo {pages} {247} (\bibinfo {year} {2022})},\ \Eprint {https://arxiv.org/abs/2204.05083} {arXiv:2204.05083 [hep-ph]} \BibitemShut {NoStop}%
\bibitem [{\citenamefont {McLerran}\ and\ \citenamefont {Pisarski}(2007)}]{McLerran:2007qj}%
  \BibitemOpen
  \bibfield  {author} {\bibinfo {author} {\bibfnamefont {L.}~\bibnamefont {McLerran}}\ and\ \bibinfo {author} {\bibfnamefont {R.~D.}\ \bibnamefont {Pisarski}},\ }\bibfield  {title} {\bibinfo {title} {{Phases of cold, dense quarks at large N(c)}},\ }\href {https://doi.org/10.1016/j.nuclphysa.2007.08.013} {\bibfield  {journal} {\bibinfo  {journal} {Nucl. Phys. A}\ }\textbf {\bibinfo {volume} {796}},\ \bibinfo {pages} {83} (\bibinfo {year} {2007})},\ \Eprint {https://arxiv.org/abs/0706.2191} {arXiv:0706.2191 [hep-ph]} \BibitemShut {NoStop}%
\bibitem [{\citenamefont {Detar}\ and\ \citenamefont {Kunihiro}(1989)}]{Detar:1988kn}%
  \BibitemOpen
  \bibfield  {author} {\bibinfo {author} {\bibfnamefont {C.~E.}\ \bibnamefont {Detar}}\ and\ \bibinfo {author} {\bibfnamefont {T.}~\bibnamefont {Kunihiro}},\ }\bibfield  {title} {\bibinfo {title} {{Linear $\sigma$ Model With Parity Doubling}},\ }\href {https://doi.org/10.1103/PhysRevD.39.2805} {\bibfield  {journal} {\bibinfo  {journal} {Phys. Rev. D}\ }\textbf {\bibinfo {volume} {39}},\ \bibinfo {pages} {2805} (\bibinfo {year} {1989})}\BibitemShut {NoStop}%
\bibitem [{\citenamefont {Jido}\ \emph {et~al.}(2000{\natexlab{a}})\citenamefont {Jido}, \citenamefont {Nemoto}, \citenamefont {Oka},\ and\ \citenamefont {Hosaka}}]{Jido:1998av}%
  \BibitemOpen
  \bibfield  {author} {\bibinfo {author} {\bibfnamefont {D.}~\bibnamefont {Jido}}, \bibinfo {author} {\bibfnamefont {Y.}~\bibnamefont {Nemoto}}, \bibinfo {author} {\bibfnamefont {M.}~\bibnamefont {Oka}},\ and\ \bibinfo {author} {\bibfnamefont {A.}~\bibnamefont {Hosaka}},\ }\bibfield  {title} {\bibinfo {title} {{Chiral symmetry for positive and negative parity nucleons}},\ }\href {https://doi.org/10.1016/S0375-9474(99)00844-1} {\bibfield  {journal} {\bibinfo  {journal} {Nucl. Phys. A}\ }\textbf {\bibinfo {volume} {671}},\ \bibinfo {pages} {471} (\bibinfo {year} {2000}{\natexlab{a}})},\ \Eprint {https://arxiv.org/abs/hep-ph/9805306} {arXiv:hep-ph/9805306} \BibitemShut {NoStop}%
\bibitem [{\citenamefont {Jido}\ \emph {et~al.}(2000{\natexlab{b}})\citenamefont {Jido}, \citenamefont {Hatsuda},\ and\ \citenamefont {Kunihiro}}]{Jido:1999hd}%
  \BibitemOpen
  \bibfield  {author} {\bibinfo {author} {\bibfnamefont {D.}~\bibnamefont {Jido}}, \bibinfo {author} {\bibfnamefont {T.}~\bibnamefont {Hatsuda}},\ and\ \bibinfo {author} {\bibfnamefont {T.}~\bibnamefont {Kunihiro}},\ }\bibfield  {title} {\bibinfo {title} {{Chiral symmetry realization for even parity and odd parity baryon resonances}},\ }\href {https://doi.org/10.1103/PhysRevLett.84.3252} {\bibfield  {journal} {\bibinfo  {journal} {Phys. Rev. Lett.}\ }\textbf {\bibinfo {volume} {84}},\ \bibinfo {pages} {3252} (\bibinfo {year} {2000}{\natexlab{b}})},\ \Eprint {https://arxiv.org/abs/hep-ph/9910375} {arXiv:hep-ph/9910375} \BibitemShut {NoStop}%
\bibitem [{\citenamefont {Jido}\ \emph {et~al.}(2001)\citenamefont {Jido}, \citenamefont {Oka},\ and\ \citenamefont {Hosaka}}]{Jido:2001nt}%
  \BibitemOpen
  \bibfield  {author} {\bibinfo {author} {\bibfnamefont {D.}~\bibnamefont {Jido}}, \bibinfo {author} {\bibfnamefont {M.}~\bibnamefont {Oka}},\ and\ \bibinfo {author} {\bibfnamefont {A.}~\bibnamefont {Hosaka}},\ }\bibfield  {title} {\bibinfo {title} {{Chiral symmetry of baryons}},\ }\href {https://doi.org/10.1143/PTP.106.873} {\bibfield  {journal} {\bibinfo  {journal} {Prog. Theor. Phys.}\ }\textbf {\bibinfo {volume} {106}},\ \bibinfo {pages} {873} (\bibinfo {year} {2001})},\ \Eprint {https://arxiv.org/abs/hep-ph/0110005} {arXiv:hep-ph/0110005} \BibitemShut {NoStop}%
\bibitem [{\citenamefont {Zschiesche}\ \emph {et~al.}(2007)\citenamefont {Zschiesche}, \citenamefont {Tolos}, \citenamefont {Schaffner-Bielich},\ and\ \citenamefont {Pisarski}}]{Zschiesche:2006zj}%
  \BibitemOpen
  \bibfield  {author} {\bibinfo {author} {\bibfnamefont {D.}~\bibnamefont {Zschiesche}}, \bibinfo {author} {\bibfnamefont {L.}~\bibnamefont {Tolos}}, \bibinfo {author} {\bibfnamefont {J.}~\bibnamefont {Schaffner-Bielich}},\ and\ \bibinfo {author} {\bibfnamefont {R.~D.}\ \bibnamefont {Pisarski}},\ }\bibfield  {title} {\bibinfo {title} {{Cold, dense nuclear matter in a SU(2) parity doublet model}},\ }\href {https://doi.org/10.1103/PhysRevC.75.055202} {\bibfield  {journal} {\bibinfo  {journal} {Phys. Rev. C}\ }\textbf {\bibinfo {volume} {75}},\ \bibinfo {pages} {055202} (\bibinfo {year} {2007})},\ \Eprint {https://arxiv.org/abs/nucl-th/0608044} {arXiv:nucl-th/0608044} \BibitemShut {NoStop}%
\bibitem [{\citenamefont {Dexheimer}\ \emph {et~al.}(2008{\natexlab{a}})\citenamefont {Dexheimer}, \citenamefont {Schramm},\ and\ \citenamefont {Zschiesche}}]{Dexheimer:2007tn}%
  \BibitemOpen
  \bibfield  {author} {\bibinfo {author} {\bibfnamefont {V.}~\bibnamefont {Dexheimer}}, \bibinfo {author} {\bibfnamefont {S.}~\bibnamefont {Schramm}},\ and\ \bibinfo {author} {\bibfnamefont {D.}~\bibnamefont {Zschiesche}},\ }\bibfield  {title} {\bibinfo {title} {{Nuclear matter and neutron stars in a parity doublet model}},\ }\href {https://doi.org/10.1103/PhysRevC.77.025803} {\bibfield  {journal} {\bibinfo  {journal} {Phys. Rev. C}\ }\textbf {\bibinfo {volume} {77}},\ \bibinfo {pages} {025803} (\bibinfo {year} {2008}{\natexlab{a}})},\ \Eprint {https://arxiv.org/abs/0710.4192} {arXiv:0710.4192 [nucl-th]} \BibitemShut {NoStop}%
\bibitem [{\citenamefont {Dexheimer}\ \emph {et~al.}(2008{\natexlab{b}})\citenamefont {Dexheimer}, \citenamefont {Pagliara}, \citenamefont {Tolos}, \citenamefont {Schaffner-Bielich},\ and\ \citenamefont {Schramm}}]{Dexheimer:2008cv}%
  \BibitemOpen
  \bibfield  {author} {\bibinfo {author} {\bibfnamefont {V.}~\bibnamefont {Dexheimer}}, \bibinfo {author} {\bibfnamefont {G.}~\bibnamefont {Pagliara}}, \bibinfo {author} {\bibfnamefont {L.}~\bibnamefont {Tolos}}, \bibinfo {author} {\bibfnamefont {J.}~\bibnamefont {Schaffner-Bielich}},\ and\ \bibinfo {author} {\bibfnamefont {S.}~\bibnamefont {Schramm}},\ }\bibfield  {title} {\bibinfo {title} {{Neutron stars within the SU(2) parity doublet model}},\ }\href {https://doi.org/10.1140/epja/i2008-10652-0} {\bibfield  {journal} {\bibinfo  {journal} {Eur. Phys. J. A}\ }\textbf {\bibinfo {volume} {38}},\ \bibinfo {pages} {105} (\bibinfo {year} {2008}{\natexlab{b}})},\ \Eprint {https://arxiv.org/abs/0805.3301} {arXiv:0805.3301 [nucl-th]} \BibitemShut {NoStop}%
\bibitem [{\citenamefont {Gallas}\ \emph {et~al.}(2010)\citenamefont {Gallas}, \citenamefont {Giacosa},\ and\ \citenamefont {Rischke}}]{Gallas:2009qp}%
  \BibitemOpen
  \bibfield  {author} {\bibinfo {author} {\bibfnamefont {S.}~\bibnamefont {Gallas}}, \bibinfo {author} {\bibfnamefont {F.}~\bibnamefont {Giacosa}},\ and\ \bibinfo {author} {\bibfnamefont {D.~H.}\ \bibnamefont {Rischke}},\ }\bibfield  {title} {\bibinfo {title} {{Vacuum phenomenology of the chiral partner of the nucleon in a linear sigma model with vector mesons}},\ }\href {https://doi.org/10.1103/PhysRevD.82.014004} {\bibfield  {journal} {\bibinfo  {journal} {Phys. Rev. D}\ }\textbf {\bibinfo {volume} {82}},\ \bibinfo {pages} {014004} (\bibinfo {year} {2010})},\ \Eprint {https://arxiv.org/abs/0907.5084} {arXiv:0907.5084 [hep-ph]} \BibitemShut {NoStop}%
\bibitem [{\citenamefont {Parganlija}\ \emph {et~al.}(2010)\citenamefont {Parganlija}, \citenamefont {Giacosa},\ and\ \citenamefont {Rischke}}]{Parganlija:2010fz}%
  \BibitemOpen
  \bibfield  {author} {\bibinfo {author} {\bibfnamefont {D.}~\bibnamefont {Parganlija}}, \bibinfo {author} {\bibfnamefont {F.}~\bibnamefont {Giacosa}},\ and\ \bibinfo {author} {\bibfnamefont {D.~H.}\ \bibnamefont {Rischke}},\ }\bibfield  {title} {\bibinfo {title} {{Vacuum Properties of Mesons in a Linear Sigma Model with Vector Mesons and Global Chiral Invariance}},\ }\href {https://doi.org/10.1103/PhysRevD.82.054024} {\bibfield  {journal} {\bibinfo  {journal} {Phys. Rev. D}\ }\textbf {\bibinfo {volume} {82}},\ \bibinfo {pages} {054024} (\bibinfo {year} {2010})},\ \Eprint {https://arxiv.org/abs/1003.4934} {arXiv:1003.4934 [hep-ph]} \BibitemShut {NoStop}%
\bibitem [{\citenamefont {Gallas}\ \emph {et~al.}(2011)\citenamefont {Gallas}, \citenamefont {Giacosa},\ and\ \citenamefont {Pagliara}}]{Gallas:2011qp}%
  \BibitemOpen
  \bibfield  {author} {\bibinfo {author} {\bibfnamefont {S.}~\bibnamefont {Gallas}}, \bibinfo {author} {\bibfnamefont {F.}~\bibnamefont {Giacosa}},\ and\ \bibinfo {author} {\bibfnamefont {G.}~\bibnamefont {Pagliara}},\ }\bibfield  {title} {\bibinfo {title} {{Nuclear matter within a dilatation-invariant parity doublet model: the role of the tetraquark at nonzero density}},\ }\href {https://doi.org/10.1016/j.nuclphysa.2011.09.008} {\bibfield  {journal} {\bibinfo  {journal} {Nucl. Phys. A}\ }\textbf {\bibinfo {volume} {872}},\ \bibinfo {pages} {13} (\bibinfo {year} {2011})},\ \Eprint {https://arxiv.org/abs/1105.5003} {arXiv:1105.5003 [hep-ph]} \BibitemShut {NoStop}%
\bibitem [{\citenamefont {Steinheimer}\ \emph {et~al.}(2011)\citenamefont {Steinheimer}, \citenamefont {Schramm},\ and\ \citenamefont {Stocker}}]{Steinheimer:2011ea}%
  \BibitemOpen
  \bibfield  {author} {\bibinfo {author} {\bibfnamefont {J.}~\bibnamefont {Steinheimer}}, \bibinfo {author} {\bibfnamefont {S.}~\bibnamefont {Schramm}},\ and\ \bibinfo {author} {\bibfnamefont {H.}~\bibnamefont {Stocker}},\ }\bibfield  {title} {\bibinfo {title} {{The hadronic SU(3) Parity Doublet Model for Dense Matter, its extension to quarks and the strange equation of state}},\ }\href {https://doi.org/10.1103/PhysRevC.84.045208} {\bibfield  {journal} {\bibinfo  {journal} {Phys. Rev. C}\ }\textbf {\bibinfo {volume} {84}},\ \bibinfo {pages} {045208} (\bibinfo {year} {2011})},\ \Eprint {https://arxiv.org/abs/1108.2596} {arXiv:1108.2596 [hep-ph]} \BibitemShut {NoStop}%
\bibitem [{\citenamefont {Dexheimer}\ \emph {et~al.}(2013)\citenamefont {Dexheimer}, \citenamefont {Steinheimer}, \citenamefont {Negreiros},\ and\ \citenamefont {Schramm}}]{Dexheimer:2012eu}%
  \BibitemOpen
  \bibfield  {author} {\bibinfo {author} {\bibfnamefont {V.}~\bibnamefont {Dexheimer}}, \bibinfo {author} {\bibfnamefont {J.}~\bibnamefont {Steinheimer}}, \bibinfo {author} {\bibfnamefont {R.}~\bibnamefont {Negreiros}},\ and\ \bibinfo {author} {\bibfnamefont {S.}~\bibnamefont {Schramm}},\ }\bibfield  {title} {\bibinfo {title} {{Hybrid Stars in an SU(3) parity doublet model}},\ }\href {https://doi.org/10.1103/PhysRevC.87.015804} {\bibfield  {journal} {\bibinfo  {journal} {Phys. Rev. C}\ }\textbf {\bibinfo {volume} {87}},\ \bibinfo {pages} {015804} (\bibinfo {year} {2013})},\ \Eprint {https://arxiv.org/abs/1206.3086} {arXiv:1206.3086 [astro-ph.HE]} \BibitemShut {NoStop}%
\bibitem [{\citenamefont {Parganlija}\ \emph {et~al.}(2013)\citenamefont {Parganlija}, \citenamefont {Kovacs}, \citenamefont {Wolf}, \citenamefont {Giacosa},\ and\ \citenamefont {Rischke}}]{Parganlija:2012fy}%
  \BibitemOpen
  \bibfield  {author} {\bibinfo {author} {\bibfnamefont {D.}~\bibnamefont {Parganlija}}, \bibinfo {author} {\bibfnamefont {P.}~\bibnamefont {Kovacs}}, \bibinfo {author} {\bibfnamefont {G.}~\bibnamefont {Wolf}}, \bibinfo {author} {\bibfnamefont {F.}~\bibnamefont {Giacosa}},\ and\ \bibinfo {author} {\bibfnamefont {D.~H.}\ \bibnamefont {Rischke}},\ }\bibfield  {title} {\bibinfo {title} {{Meson vacuum phenomenology in a three-flavor linear sigma model with (axial-)vector mesons}},\ }\href {https://doi.org/10.1103/PhysRevD.87.014011} {\bibfield  {journal} {\bibinfo  {journal} {Phys. Rev. D}\ }\textbf {\bibinfo {volume} {87}},\ \bibinfo {pages} {014011} (\bibinfo {year} {2013})},\ \Eprint {https://arxiv.org/abs/1208.0585} {arXiv:1208.0585 [hep-ph]} \BibitemShut {NoStop}%
\bibitem [{\citenamefont {Holt}\ \emph {et~al.}(2016)\citenamefont {Holt}, \citenamefont {Rho},\ and\ \citenamefont {Weise}}]{Holt:2014hma}%
  \BibitemOpen
  \bibfield  {author} {\bibinfo {author} {\bibfnamefont {J.~W.}\ \bibnamefont {Holt}}, \bibinfo {author} {\bibfnamefont {M.}~\bibnamefont {Rho}},\ and\ \bibinfo {author} {\bibfnamefont {W.}~\bibnamefont {Weise}},\ }\bibfield  {title} {\bibinfo {title} {{Chiral symmetry and effective field theories for hadronic, nuclear and stellar matter}},\ }\href {https://doi.org/10.1016/j.physrep.2015.10.011} {\bibfield  {journal} {\bibinfo  {journal} {Phys. Rept.}\ }\textbf {\bibinfo {volume} {621}},\ \bibinfo {pages} {2} (\bibinfo {year} {2016})},\ \Eprint {https://arxiv.org/abs/1411.6681} {arXiv:1411.6681 [nucl-th]} \BibitemShut {NoStop}%
\bibitem [{\citenamefont {Aarts}\ \emph {et~al.}(2017)\citenamefont {Aarts}, \citenamefont {Allton}, \citenamefont {De~Boni}, \citenamefont {Hands}, \citenamefont {J{\"a}ger}, \citenamefont {Praki},\ and\ \citenamefont {Skullerud}}]{Aarts:2017rrl}%
  \BibitemOpen
  \bibfield  {author} {\bibinfo {author} {\bibfnamefont {G.}~\bibnamefont {Aarts}}, \bibinfo {author} {\bibfnamefont {C.}~\bibnamefont {Allton}}, \bibinfo {author} {\bibfnamefont {D.}~\bibnamefont {De~Boni}}, \bibinfo {author} {\bibfnamefont {S.}~\bibnamefont {Hands}}, \bibinfo {author} {\bibfnamefont {B.}~\bibnamefont {J{\"a}ger}}, \bibinfo {author} {\bibfnamefont {C.}~\bibnamefont {Praki}},\ and\ \bibinfo {author} {\bibfnamefont {J.-I.}\ \bibnamefont {Skullerud}},\ }\bibfield  {title} {\bibinfo {title} {{Light baryons below and above the deconfinement transition: medium effects and parity doubling}},\ }\href {https://doi.org/10.1007/JHEP06(2017)034} {\bibfield  {journal} {\bibinfo  {journal} {JHEP}\ }\textbf {\bibinfo {volume} {06}},\ \bibinfo {pages} {034}},\ \Eprint {https://arxiv.org/abs/1703.09246} {arXiv:1703.09246 [hep-lat]} \BibitemShut {NoStop}%
\bibitem [{\citenamefont {Mukherjee}\ \emph {et~al.}(2017)\citenamefont {Mukherjee}, \citenamefont {Schramm}, \citenamefont {Steinheimer},\ and\ \citenamefont {Dexheimer}}]{Mukherjee:2017jzi}%
  \BibitemOpen
  \bibfield  {author} {\bibinfo {author} {\bibfnamefont {A.}~\bibnamefont {Mukherjee}}, \bibinfo {author} {\bibfnamefont {S.}~\bibnamefont {Schramm}}, \bibinfo {author} {\bibfnamefont {J.}~\bibnamefont {Steinheimer}},\ and\ \bibinfo {author} {\bibfnamefont {V.}~\bibnamefont {Dexheimer}},\ }\bibfield  {title} {\bibinfo {title} {{The application of the Quark-Hadron Chiral Parity-Doublet Model to neutron star matter}},\ }\href {https://doi.org/10.1051/0004-6361/201731505} {\bibfield  {journal} {\bibinfo  {journal} {Astron. Astrophys.}\ }\textbf {\bibinfo {volume} {608}},\ \bibinfo {pages} {A110} (\bibinfo {year} {2017})},\ \Eprint {https://arxiv.org/abs/1706.09191} {arXiv:1706.09191 [nucl-th]} \BibitemShut {NoStop}%
\bibitem [{\citenamefont {Catillo}\ and\ \citenamefont {Glozman}(2018)}]{Catillo:2018cyv}%
  \BibitemOpen
  \bibfield  {author} {\bibinfo {author} {\bibfnamefont {M.}~\bibnamefont {Catillo}}\ and\ \bibinfo {author} {\bibfnamefont {L.~Y.}\ \bibnamefont {Glozman}},\ }\bibfield  {title} {\bibinfo {title} {{Baryon parity doublets and chiral spin symmetry}},\ }\href {https://doi.org/10.1103/PhysRevD.98.014030} {\bibfield  {journal} {\bibinfo  {journal} {Phys. Rev. D}\ }\textbf {\bibinfo {volume} {98}},\ \bibinfo {pages} {014030} (\bibinfo {year} {2018})},\ \Eprint {https://arxiv.org/abs/1804.07171} {arXiv:1804.07171 [hep-ph]} \BibitemShut {NoStop}%
\bibitem [{\citenamefont {Marczenko}\ \emph {et~al.}(2018)\citenamefont {Marczenko}, \citenamefont {Blaschke}, \citenamefont {Redlich},\ and\ \citenamefont {Sasaki}}]{Marczenko:2018jui}%
  \BibitemOpen
  \bibfield  {author} {\bibinfo {author} {\bibfnamefont {M.}~\bibnamefont {Marczenko}}, \bibinfo {author} {\bibfnamefont {D.}~\bibnamefont {Blaschke}}, \bibinfo {author} {\bibfnamefont {K.}~\bibnamefont {Redlich}},\ and\ \bibinfo {author} {\bibfnamefont {C.}~\bibnamefont {Sasaki}},\ }\bibfield  {title} {\bibinfo {title} {{Chiral symmetry restoration by parity doubling and the structure of neutron stars}},\ }\href {https://doi.org/10.1103/PhysRevD.98.103021} {\bibfield  {journal} {\bibinfo  {journal} {Phys. Rev. D}\ }\textbf {\bibinfo {volume} {98}},\ \bibinfo {pages} {103021} (\bibinfo {year} {2018})},\ \Eprint {https://arxiv.org/abs/1805.06886} {arXiv:1805.06886 [nucl-th]} \BibitemShut {NoStop}%
\bibitem [{\citenamefont {Motornenko}\ \emph {et~al.}(2020)\citenamefont {Motornenko}, \citenamefont {Steinheimer}, \citenamefont {Vovchenko}, \citenamefont {Schramm},\ and\ \citenamefont {Stoecker}}]{Motornenko:2019arp}%
  \BibitemOpen
  \bibfield  {author} {\bibinfo {author} {\bibfnamefont {A.}~\bibnamefont {Motornenko}}, \bibinfo {author} {\bibfnamefont {J.}~\bibnamefont {Steinheimer}}, \bibinfo {author} {\bibfnamefont {V.}~\bibnamefont {Vovchenko}}, \bibinfo {author} {\bibfnamefont {S.}~\bibnamefont {Schramm}},\ and\ \bibinfo {author} {\bibfnamefont {H.}~\bibnamefont {Stoecker}},\ }\bibfield  {title} {\bibinfo {title} {{Equation of state for hot QCD and compact stars from a mean field approach}},\ }\href {https://doi.org/10.1103/PhysRevC.101.034904} {\bibfield  {journal} {\bibinfo  {journal} {Phys. Rev. C}\ }\textbf {\bibinfo {volume} {101}},\ \bibinfo {pages} {034904} (\bibinfo {year} {2020})},\ \Eprint {https://arxiv.org/abs/1905.00866} {arXiv:1905.00866 [hep-ph]} \BibitemShut {NoStop}%
\bibitem [{\citenamefont {Marczenko}\ \emph {et~al.}(2019)\citenamefont {Marczenko}, \citenamefont {Blaschke}, \citenamefont {Redlich},\ and\ \citenamefont {Sasaki}}]{Marczenko:2019trv}%
  \BibitemOpen
  \bibfield  {author} {\bibinfo {author} {\bibfnamefont {M.}~\bibnamefont {Marczenko}}, \bibinfo {author} {\bibfnamefont {D.}~\bibnamefont {Blaschke}}, \bibinfo {author} {\bibfnamefont {K.}~\bibnamefont {Redlich}},\ and\ \bibinfo {author} {\bibfnamefont {C.}~\bibnamefont {Sasaki}},\ }\bibfield  {title} {\bibinfo {title} {{Parity Doubling and the Dense Matter Phase Diagram under Constraints from Multi-Messenger Astronomy}},\ }\href {https://doi.org/10.3390/universe5080180} {\bibfield  {journal} {\bibinfo  {journal} {Universe}\ }\textbf {\bibinfo {volume} {5}},\ \bibinfo {pages} {180} (\bibinfo {year} {2019})},\ \Eprint {https://arxiv.org/abs/1905.04974} {arXiv:1905.04974 [nucl-th]} \BibitemShut {NoStop}%
\bibitem [{\citenamefont {Marczenko}\ \emph {et~al.}(2020)\citenamefont {Marczenko}, \citenamefont {Blaschke}, \citenamefont {Redlich},\ and\ \citenamefont {Sasaki}}]{Marczenko:2020jma}%
  \BibitemOpen
  \bibfield  {author} {\bibinfo {author} {\bibfnamefont {M.}~\bibnamefont {Marczenko}}, \bibinfo {author} {\bibfnamefont {D.}~\bibnamefont {Blaschke}}, \bibinfo {author} {\bibfnamefont {K.}~\bibnamefont {Redlich}},\ and\ \bibinfo {author} {\bibfnamefont {C.}~\bibnamefont {Sasaki}},\ }\bibfield  {title} {\bibinfo {title} {{Toward a unified equation of state for multi-messenger astronomy}},\ }\href {https://doi.org/10.1051/0004-6361/202038211} {\bibfield  {journal} {\bibinfo  {journal} {Astron. Astrophys.}\ }\textbf {\bibinfo {volume} {643}},\ \bibinfo {pages} {A82} (\bibinfo {year} {2020})},\ \Eprint {https://arxiv.org/abs/2004.09566} {arXiv:2004.09566 [astro-ph.HE]} \BibitemShut {NoStop}%
\bibitem [{\citenamefont {Minamikawa}\ \emph {et~al.}(2021{\natexlab{a}})\citenamefont {Minamikawa}, \citenamefont {Kojo},\ and\ \citenamefont {Harada}}]{Minamikawa:2020jfj}%
  \BibitemOpen
  \bibfield  {author} {\bibinfo {author} {\bibfnamefont {T.}~\bibnamefont {Minamikawa}}, \bibinfo {author} {\bibfnamefont {T.}~\bibnamefont {Kojo}},\ and\ \bibinfo {author} {\bibfnamefont {M.}~\bibnamefont {Harada}},\ }\bibfield  {title} {\bibinfo {title} {{Quark-hadron crossover equations of state for neutron stars: constraining the chiral invariant mass in a parity doublet model}},\ }\href {https://doi.org/10.1103/PhysRevC.103.045205} {\bibfield  {journal} {\bibinfo  {journal} {Phys. Rev. C}\ }\textbf {\bibinfo {volume} {103}},\ \bibinfo {pages} {045205} (\bibinfo {year} {2021}{\natexlab{a}})},\ \Eprint {https://arxiv.org/abs/2011.13684} {arXiv:2011.13684 [nucl-th]} \BibitemShut {NoStop}%
\bibitem [{\citenamefont {Pisarski}(2021)}]{Pisarski:2021aoz}%
  \BibitemOpen
  \bibfield  {author} {\bibinfo {author} {\bibfnamefont {R.~D.}\ \bibnamefont {Pisarski}},\ }\bibfield  {title} {\bibinfo {title} {{Remarks on nuclear matter: How an $\omega_0$ condensate can spike the speed of sound, and a model of $Z(3)$ baryons}},\ }\href {https://doi.org/10.1103/PhysRevD.103.L071504} {\bibfield  {journal} {\bibinfo  {journal} {Phys. Rev. D}\ }\textbf {\bibinfo {volume} {103}},\ \bibinfo {pages} {L071504} (\bibinfo {year} {2021})},\ \Eprint {https://arxiv.org/abs/2101.05813} {arXiv:2101.05813 [nucl-th]} \BibitemShut {NoStop}%
\bibitem [{\citenamefont {Minamikawa}\ \emph {et~al.}(2021{\natexlab{b}})\citenamefont {Minamikawa}, \citenamefont {Kojo},\ and\ \citenamefont {Harada}}]{Minamikawa:2021fln}%
  \BibitemOpen
  \bibfield  {author} {\bibinfo {author} {\bibfnamefont {T.}~\bibnamefont {Minamikawa}}, \bibinfo {author} {\bibfnamefont {T.}~\bibnamefont {Kojo}},\ and\ \bibinfo {author} {\bibfnamefont {M.}~\bibnamefont {Harada}},\ }\bibfield  {title} {\bibinfo {title} {{Chiral condensates for neutron stars in hadron-quark crossover: From a parity doublet nucleon model to a Nambu\textendash{}Jona-Lasinio quark model}},\ }\href {https://doi.org/10.1103/PhysRevC.104.065201} {\bibfield  {journal} {\bibinfo  {journal} {Phys. Rev. C}\ }\textbf {\bibinfo {volume} {104}},\ \bibinfo {pages} {065201} (\bibinfo {year} {2021}{\natexlab{b}})},\ \Eprint {https://arxiv.org/abs/2107.14545} {arXiv:2107.14545 [nucl-th]} \BibitemShut {NoStop}%
\bibitem [{\citenamefont {Marczenko}\ \emph {et~al.}(2022{\natexlab{a}})\citenamefont {Marczenko}, \citenamefont {Redlich},\ and\ \citenamefont {Sasaki}}]{Marczenko:2021uaj}%
  \BibitemOpen
  \bibfield  {author} {\bibinfo {author} {\bibfnamefont {M.}~\bibnamefont {Marczenko}}, \bibinfo {author} {\bibfnamefont {K.}~\bibnamefont {Redlich}},\ and\ \bibinfo {author} {\bibfnamefont {C.}~\bibnamefont {Sasaki}},\ }\bibfield  {title} {\bibinfo {title} {{Reconciling Multi-messenger Constraints with Chiral Symmetry Restoration}},\ }\href {https://doi.org/10.3847/2041-8213/ac4b61} {\bibfield  {journal} {\bibinfo  {journal} {Astrophys. J. Lett.}\ }\textbf {\bibinfo {volume} {925}},\ \bibinfo {pages} {L23} (\bibinfo {year} {2022}{\natexlab{a}})},\ \Eprint {https://arxiv.org/abs/2110.11056} {arXiv:2110.11056 [nucl-th]} \BibitemShut {NoStop}%
\bibitem [{\citenamefont {Marczenko}\ \emph {et~al.}(2022{\natexlab{b}})\citenamefont {Marczenko}, \citenamefont {Redlich},\ and\ \citenamefont {Sasaki}}]{Marczenko:2022hyt}%
  \BibitemOpen
  \bibfield  {author} {\bibinfo {author} {\bibfnamefont {M.}~\bibnamefont {Marczenko}}, \bibinfo {author} {\bibfnamefont {K.}~\bibnamefont {Redlich}},\ and\ \bibinfo {author} {\bibfnamefont {C.}~\bibnamefont {Sasaki}},\ }\bibfield  {title} {\bibinfo {title} {{Chiral symmetry restoration and \ensuremath{\Delta} matter formation in neutron stars}},\ }\href {https://doi.org/10.1103/PhysRevD.105.103009} {\bibfield  {journal} {\bibinfo  {journal} {Phys. Rev. D}\ }\textbf {\bibinfo {volume} {105}},\ \bibinfo {pages} {103009} (\bibinfo {year} {2022}{\natexlab{b}})},\ \Eprint {https://arxiv.org/abs/2203.00269} {arXiv:2203.00269 [nucl-th]} \BibitemShut {NoStop}%
\bibitem [{\citenamefont {Gao}\ \emph {et~al.}(2022)\citenamefont {Gao}, \citenamefont {Minamikawa}, \citenamefont {Kojo},\ and\ \citenamefont {Harada}}]{Gao:2022klm}%
  \BibitemOpen
  \bibfield  {author} {\bibinfo {author} {\bibfnamefont {B.}~\bibnamefont {Gao}}, \bibinfo {author} {\bibfnamefont {T.}~\bibnamefont {Minamikawa}}, \bibinfo {author} {\bibfnamefont {T.}~\bibnamefont {Kojo}},\ and\ \bibinfo {author} {\bibfnamefont {M.}~\bibnamefont {Harada}},\ }\bibfield  {title} {\bibinfo {title} {{Impacts of the U(1)A anomaly on nuclear and neutron star equation~of state based on a parity doublet model}},\ }\href {https://doi.org/10.1103/PhysRevC.106.065205} {\bibfield  {journal} {\bibinfo  {journal} {Phys. Rev. C}\ }\textbf {\bibinfo {volume} {106}},\ \bibinfo {pages} {065205} (\bibinfo {year} {2022})},\ \Eprint {https://arxiv.org/abs/2207.05970} {arXiv:2207.05970 [nucl-th]} \BibitemShut {NoStop}%
\bibitem [{\citenamefont {Minamikawa}\ \emph {et~al.}(2023{\natexlab{a}})\citenamefont {Minamikawa}, \citenamefont {Gao}, \citenamefont {Kojo},\ and\ \citenamefont {Harada}}]{Minamikawa:2023eky}%
  \BibitemOpen
  \bibfield  {author} {\bibinfo {author} {\bibfnamefont {T.}~\bibnamefont {Minamikawa}}, \bibinfo {author} {\bibfnamefont {B.}~\bibnamefont {Gao}}, \bibinfo {author} {\bibfnamefont {T.}~\bibnamefont {Kojo}},\ and\ \bibinfo {author} {\bibfnamefont {M.}~\bibnamefont {Harada}},\ }\bibfield  {title} {\bibinfo {title} {{Chiral Restoration of Nucleons in Neutron Star Matter: Studies Based on a Parity Doublet Model}},\ }\href {https://doi.org/10.3390/sym15030745} {\bibfield  {journal} {\bibinfo  {journal} {Symmetry}\ }\textbf {\bibinfo {volume} {15}},\ \bibinfo {pages} {745} (\bibinfo {year} {2023}{\natexlab{a}})},\ \Eprint {https://arxiv.org/abs/2302.00825} {arXiv:2302.00825 [nucl-th]} \BibitemShut {NoStop}%
\bibitem [{\citenamefont {Kong}\ \emph {et~al.}(2023)\citenamefont {Kong}, \citenamefont {Minamikawa},\ and\ \citenamefont {Harada}}]{Kong:2023nue}%
  \BibitemOpen
  \bibfield  {author} {\bibinfo {author} {\bibfnamefont {Y.~K.}\ \bibnamefont {Kong}}, \bibinfo {author} {\bibfnamefont {T.}~\bibnamefont {Minamikawa}},\ and\ \bibinfo {author} {\bibfnamefont {M.}~\bibnamefont {Harada}},\ }\bibfield  {title} {\bibinfo {title} {{Neutron star matter based on a parity doublet model including the a0(980) meson}},\ }\href {https://doi.org/10.1103/PhysRevC.108.055206} {\bibfield  {journal} {\bibinfo  {journal} {Phys. Rev. C}\ }\textbf {\bibinfo {volume} {108}},\ \bibinfo {pages} {055206} (\bibinfo {year} {2023})},\ \Eprint {https://arxiv.org/abs/2306.08140} {arXiv:2306.08140 [nucl-th]} \BibitemShut {NoStop}%
\bibitem [{\citenamefont {Minamikawa}\ \emph {et~al.}(2023{\natexlab{b}})\citenamefont {Minamikawa}, \citenamefont {Gao}, \citenamefont {kojo},\ and\ \citenamefont {Harada}}]{Minamikawa:2023ypn}%
  \BibitemOpen
  \bibfield  {author} {\bibinfo {author} {\bibfnamefont {T.}~\bibnamefont {Minamikawa}}, \bibinfo {author} {\bibfnamefont {B.}~\bibnamefont {Gao}}, \bibinfo {author} {\bibfnamefont {T.}~\bibnamefont {kojo}},\ and\ \bibinfo {author} {\bibfnamefont {M.}~\bibnamefont {Harada}},\ }\bibfield  {title} {\bibinfo {title} {{Parity doublet model for baryon octets: Diquark classifications and mass hierarchy based on the quark-line diagram}},\ }\href {https://doi.org/10.1103/PhysRevD.108.076017} {\bibfield  {journal} {\bibinfo  {journal} {Phys. Rev. D}\ }\textbf {\bibinfo {volume} {108}},\ \bibinfo {pages} {076017} (\bibinfo {year} {2023}{\natexlab{b}})},\ \Eprint {https://arxiv.org/abs/2306.15564} {arXiv:2306.15564 [hep-ph]} \BibitemShut {NoStop}%
\bibitem [{\citenamefont {Koch}\ \emph {et~al.}(2024)\citenamefont {Koch}, \citenamefont {Marczenko}, \citenamefont {Redlich},\ and\ \citenamefont {Sasaki}}]{Koch:2023oez}%
  \BibitemOpen
  \bibfield  {author} {\bibinfo {author} {\bibfnamefont {V.}~\bibnamefont {Koch}}, \bibinfo {author} {\bibfnamefont {M.}~\bibnamefont {Marczenko}}, \bibinfo {author} {\bibfnamefont {K.}~\bibnamefont {Redlich}},\ and\ \bibinfo {author} {\bibfnamefont {C.}~\bibnamefont {Sasaki}},\ }\bibfield  {title} {\bibinfo {title} {{Fluctuations and correlations of baryonic chiral partners}},\ }\href {https://doi.org/10.1103/PhysRevD.109.014033} {\bibfield  {journal} {\bibinfo  {journal} {Phys. Rev. D}\ }\textbf {\bibinfo {volume} {109}},\ \bibinfo {pages} {014033} (\bibinfo {year} {2024})},\ \Eprint {https://arxiv.org/abs/2308.15794} {arXiv:2308.15794 [hep-ph]} \BibitemShut {NoStop}%
\bibitem [{\citenamefont {Fraga}\ \emph {et~al.}(2023)\citenamefont {Fraga}, \citenamefont {da~Mata},\ and\ \citenamefont {Schaffner-Bielich}}]{Fraga:2023wtd}%
  \BibitemOpen
  \bibfield  {author} {\bibinfo {author} {\bibfnamefont {E.~S.}\ \bibnamefont {Fraga}}, \bibinfo {author} {\bibfnamefont {R.}~\bibnamefont {da~Mata}},\ and\ \bibinfo {author} {\bibfnamefont {J.}~\bibnamefont {Schaffner-Bielich}},\ }\bibfield  {title} {\bibinfo {title} {{SU(3) parity doubling in cold neutron star matter}},\ }\href {https://doi.org/10.1103/PhysRevD.108.116003} {\bibfield  {journal} {\bibinfo  {journal} {Phys. Rev. D}\ }\textbf {\bibinfo {volume} {108}},\ \bibinfo {pages} {116003} (\bibinfo {year} {2023})},\ \Eprint {https://arxiv.org/abs/2309.02368} {arXiv:2309.02368 [hep-ph]} \BibitemShut {NoStop}%
\bibitem [{\citenamefont {Eser}\ and\ \citenamefont {Blaizot}(2024{\natexlab{a}})}]{Eser:2023oii}%
  \BibitemOpen
  \bibfield  {author} {\bibinfo {author} {\bibfnamefont {J.}~\bibnamefont {Eser}}\ and\ \bibinfo {author} {\bibfnamefont {J.-P.}\ \bibnamefont {Blaizot}},\ }\bibfield  {title} {\bibinfo {title} {{Thermodynamics of the parity-doublet model: Symmetric nuclear matter and the chiral transition}},\ }\href {https://doi.org/10.1103/PhysRevC.109.045201} {\bibfield  {journal} {\bibinfo  {journal} {Phys. Rev. C}\ }\textbf {\bibinfo {volume} {109}},\ \bibinfo {pages} {045201} (\bibinfo {year} {2024}{\natexlab{a}})},\ \Eprint {https://arxiv.org/abs/2309.06566} {arXiv:2309.06566 [nucl-th]} \BibitemShut {NoStop}%
\bibitem [{\citenamefont {Gao}\ \emph {et~al.}(2024{\natexlab{a}})\citenamefont {Gao}, \citenamefont {Kojo},\ and\ \citenamefont {Harada}}]{Gao:2024mew}%
  \BibitemOpen
  \bibfield  {author} {\bibinfo {author} {\bibfnamefont {B.}~\bibnamefont {Gao}}, \bibinfo {author} {\bibfnamefont {T.}~\bibnamefont {Kojo}},\ and\ \bibinfo {author} {\bibfnamefont {M.}~\bibnamefont {Harada}},\ }\bibfield  {title} {\bibinfo {title} {{Parity doublet model for baryon octets: Ground states saturated by good diquarks and the role of bad diquarks for excited states}},\ }\href {https://doi.org/10.1103/PhysRevD.110.016016} {\bibfield  {journal} {\bibinfo  {journal} {Phys. Rev. D}\ }\textbf {\bibinfo {volume} {110}},\ \bibinfo {pages} {016016} (\bibinfo {year} {2024}{\natexlab{a}})},\ \Eprint {https://arxiv.org/abs/2403.18214} {arXiv:2403.18214 [hep-ph]} \BibitemShut {NoStop}%
\bibitem [{\citenamefont {Gao}\ \emph {et~al.}(2024{\natexlab{b}})\citenamefont {Gao}, \citenamefont {Yuan}, \citenamefont {Harada},\ and\ \citenamefont {Ma}}]{Gao:2024lzu}%
  \BibitemOpen
  \bibfield  {author} {\bibinfo {author} {\bibfnamefont {B.}~\bibnamefont {Gao}}, \bibinfo {author} {\bibfnamefont {W.-L.}\ \bibnamefont {Yuan}}, \bibinfo {author} {\bibfnamefont {M.}~\bibnamefont {Harada}},\ and\ \bibinfo {author} {\bibfnamefont {Y.-L.}\ \bibnamefont {Ma}},\ }\bibfield  {title} {\bibinfo {title} {{Exploring the first-order phase transition in neutron stars using the parity doublet model and a Nambu\textendash{}Jona-Lasinio\textendash{}type quark model}},\ }\href {https://doi.org/10.1103/PhysRevC.110.045802} {\bibfield  {journal} {\bibinfo  {journal} {Phys. Rev. C}\ }\textbf {\bibinfo {volume} {110}},\ \bibinfo {pages} {045802} (\bibinfo {year} {2024}{\natexlab{b}})},\ \Eprint {https://arxiv.org/abs/2407.13990} {arXiv:2407.13990 [nucl-th]} \BibitemShut {NoStop}%
\bibitem [{\citenamefont {Giacosa}\ \emph {et~al.}(2025{\natexlab{a}})\citenamefont {Giacosa}, \citenamefont {Kov{\'a}cs},\ and\ \citenamefont {Jafarzade}}]{Giacosa:2024epf}%
  \BibitemOpen
  \bibfield  {author} {\bibinfo {author} {\bibfnamefont {F.}~\bibnamefont {Giacosa}}, \bibinfo {author} {\bibfnamefont {P.}~\bibnamefont {Kov{\'a}cs}},\ and\ \bibinfo {author} {\bibfnamefont {S.}~\bibnamefont {Jafarzade}},\ }\bibfield  {title} {\bibinfo {title} {{Ordinary and exotic mesons in the extended Linear Sigma Model}},\ }\href {https://doi.org/10.1016/j.ppnp.2025.104176} {\bibfield  {journal} {\bibinfo  {journal} {Prog. Part. Nucl. Phys.}\ }\textbf {\bibinfo {volume} {143}},\ \bibinfo {pages} {104176} (\bibinfo {year} {2025}{\natexlab{a}})},\ \Eprint {https://arxiv.org/abs/2407.18348} {arXiv:2407.18348 [hep-ph]} \BibitemShut {NoStop}%
\bibitem [{\citenamefont {Eser}\ and\ \citenamefont {Blaizot}(2024{\natexlab{b}})}]{Eser:2024xil}%
  \BibitemOpen
  \bibfield  {author} {\bibinfo {author} {\bibfnamefont {J.}~\bibnamefont {Eser}}\ and\ \bibinfo {author} {\bibfnamefont {J.-P.}\ \bibnamefont {Blaizot}},\ }\bibfield  {title} {\bibinfo {title} {{Thermodynamics of the parity-doublet model. II. Asymmetric and neutron matter}},\ }\href {https://doi.org/10.1103/PhysRevC.110.065205} {\bibfield  {journal} {\bibinfo  {journal} {Phys. Rev. C}\ }\textbf {\bibinfo {volume} {110}},\ \bibinfo {pages} {065205} (\bibinfo {year} {2024}{\natexlab{b}})},\ \Eprint {https://arxiv.org/abs/2408.01302} {arXiv:2408.01302 [nucl-th]} \BibitemShut {NoStop}%
\bibitem [{\citenamefont {Yasui}\ \emph {et~al.}(2025)\citenamefont {Yasui}, \citenamefont {Nitta},\ and\ \citenamefont {Sasaki}}]{Yasui:2024dbx}%
  \BibitemOpen
  \bibfield  {author} {\bibinfo {author} {\bibfnamefont {S.}~\bibnamefont {Yasui}}, \bibinfo {author} {\bibfnamefont {M.}~\bibnamefont {Nitta}},\ and\ \bibinfo {author} {\bibfnamefont {C.}~\bibnamefont {Sasaki}},\ }\bibfield  {title} {\bibinfo {title} {{Emergent chirality and superfluidity of parity-doubled baryons in neutron stars}},\ }\href {https://doi.org/10.1103/PhysRevD.111.034029} {\bibfield  {journal} {\bibinfo  {journal} {Phys. Rev. D}\ }\textbf {\bibinfo {volume} {111}},\ \bibinfo {pages} {034029} (\bibinfo {year} {2025})},\ \Eprint {https://arxiv.org/abs/2409.05670} {arXiv:2409.05670 [nucl-th]} \BibitemShut {NoStop}%
\bibitem [{\citenamefont {Steiner}\ \emph {et~al.}(2013)\citenamefont {Steiner}, \citenamefont {Hempel},\ and\ \citenamefont {Fischer}}]{Steiner:2012rk}%
  \BibitemOpen
  \bibfield  {author} {\bibinfo {author} {\bibfnamefont {A.~W.}\ \bibnamefont {Steiner}}, \bibinfo {author} {\bibfnamefont {M.}~\bibnamefont {Hempel}},\ and\ \bibinfo {author} {\bibfnamefont {T.}~\bibnamefont {Fischer}},\ }\bibfield  {title} {\bibinfo {title} {{Core-collapse supernova equations of state based on neutron star observations}},\ }\href {https://doi.org/10.1088/0004-637X/774/1/17} {\bibfield  {journal} {\bibinfo  {journal} {Astrophys. J.}\ }\textbf {\bibinfo {volume} {774}},\ \bibinfo {pages} {17} (\bibinfo {year} {2013})},\ \Eprint {https://arxiv.org/abs/1207.2184} {arXiv:1207.2184 [astro-ph.SR]} \BibitemShut {NoStop}%
\bibitem [{\citenamefont {Alford}\ \emph {et~al.}(2023)\citenamefont {Alford}, \citenamefont {Brodie}, \citenamefont {Haber},\ and\ \citenamefont {Tews}}]{Alford:2023rgp}%
  \BibitemOpen
  \bibfield  {author} {\bibinfo {author} {\bibfnamefont {M.~G.}\ \bibnamefont {Alford}}, \bibinfo {author} {\bibfnamefont {L.}~\bibnamefont {Brodie}}, \bibinfo {author} {\bibfnamefont {A.}~\bibnamefont {Haber}},\ and\ \bibinfo {author} {\bibfnamefont {I.}~\bibnamefont {Tews}},\ }\bibfield  {title} {\bibinfo {title} {{Tabulated equations of state from models informed by chiral effective field theory}},\ }\href {https://doi.org/10.1088/1402-4896/ad03c8} {\bibfield  {journal} {\bibinfo  {journal} {Phys. Scripta}\ }\textbf {\bibinfo {volume} {98}},\ \bibinfo {pages} {125302} (\bibinfo {year} {2023})},\ \Eprint {https://arxiv.org/abs/2304.07836} {arXiv:2304.07836 [nucl-th]} \BibitemShut {NoStop}%
\bibitem [{\citenamefont {Alford}\ \emph {et~al.}(2024)\citenamefont {Alford}, \citenamefont {Haber},\ and\ \citenamefont {Zhang}}]{Alford:2024xfb}%
  \BibitemOpen
  \bibfield  {author} {\bibinfo {author} {\bibfnamefont {M.~G.}\ \bibnamefont {Alford}}, \bibinfo {author} {\bibfnamefont {A.}~\bibnamefont {Haber}},\ and\ \bibinfo {author} {\bibfnamefont {Z.}~\bibnamefont {Zhang}},\ }\bibfield  {title} {\bibinfo {title} {{Beyond modified Urca: The nucleon width approximation for flavor-changing processes~in dense matter}},\ }\href {https://doi.org/10.1103/PhysRevC.110.L052801} {\bibfield  {journal} {\bibinfo  {journal} {Phys. Rev. C}\ }\textbf {\bibinfo {volume} {110}},\ \bibinfo {pages} {L052801} (\bibinfo {year} {2024})},\ \Eprint {https://arxiv.org/abs/2406.13717} {arXiv:2406.13717 [nucl-th]} \BibitemShut {NoStop}%
\bibitem [{\citenamefont {Sedrakian}(2024)}]{Sedrakian:2024uma}%
  \BibitemOpen
  \bibfield  {author} {\bibinfo {author} {\bibfnamefont {A.}~\bibnamefont {Sedrakian}},\ }\bibfield  {title} {\bibinfo {title} {{Short-Range Correlations and Urca Process in Neutron Stars}},\ }\href {https://doi.org/10.1103/PhysRevLett.133.171401} {\bibfield  {journal} {\bibinfo  {journal} {Phys. Rev. Lett.}\ }\textbf {\bibinfo {volume} {133}},\ \bibinfo {pages} {171401} (\bibinfo {year} {2024})},\ \Eprint {https://arxiv.org/abs/2406.16183} {arXiv:2406.16183 [nucl-th]} \BibitemShut {NoStop}%
\bibitem [{\citenamefont {Pisarski}\ and\ \citenamefont {Rennecke}(2024)}]{Pisarski:2024esv}%
  \BibitemOpen
  \bibfield  {author} {\bibinfo {author} {\bibfnamefont {R.~D.}\ \bibnamefont {Pisarski}}\ and\ \bibinfo {author} {\bibfnamefont {F.}~\bibnamefont {Rennecke}},\ }\bibfield  {title} {\bibinfo {title} {{Conjectures about the Chiral Phase Transition in QCD from Anomalous Multi-Instanton Interactions}},\ }\href {https://doi.org/10.1103/PhysRevLett.132.251903} {\bibfield  {journal} {\bibinfo  {journal} {Phys. Rev. Lett.}\ }\textbf {\bibinfo {volume} {132}},\ \bibinfo {pages} {251903} (\bibinfo {year} {2024})},\ \Eprint {https://arxiv.org/abs/2401.06130} {arXiv:2401.06130 [hep-ph]} \BibitemShut {NoStop}%
\bibitem [{\citenamefont {Giacosa}\ \emph {et~al.}(2025{\natexlab{b}})\citenamefont {Giacosa}, \citenamefont {Kov{\'a}cs}, \citenamefont {Kov{\'a}cs}, \citenamefont {Pisarski},\ and\ \citenamefont {Rennecke}}]{Giacosa:2024orp}%
  \BibitemOpen
  \bibfield  {author} {\bibinfo {author} {\bibfnamefont {F.}~\bibnamefont {Giacosa}}, \bibinfo {author} {\bibfnamefont {G.}~\bibnamefont {Kov{\'a}cs}}, \bibinfo {author} {\bibfnamefont {P.}~\bibnamefont {Kov{\'a}cs}}, \bibinfo {author} {\bibfnamefont {R.~D.}\ \bibnamefont {Pisarski}},\ and\ \bibinfo {author} {\bibfnamefont {F.}~\bibnamefont {Rennecke}},\ }\bibfield  {title} {\bibinfo {title} {{Anomalous U(1)A couplings and the Columbia plot}},\ }\href {https://doi.org/10.1103/PhysRevD.111.016014} {\bibfield  {journal} {\bibinfo  {journal} {Phys. Rev. D}\ }\textbf {\bibinfo {volume} {111}},\ \bibinfo {pages} {016014} (\bibinfo {year} {2025}{\natexlab{b}})},\ \Eprint {https://arxiv.org/abs/2410.08185} {arXiv:2410.08185 [hep-ph]} \BibitemShut {NoStop}%
\bibitem [{\citenamefont {Boguta}\ and\ \citenamefont {Bodmer}(1977)}]{Boguta:1977xi}%
  \BibitemOpen
  \bibfield  {author} {\bibinfo {author} {\bibfnamefont {J.}~\bibnamefont {Boguta}}\ and\ \bibinfo {author} {\bibfnamefont {A.~R.}\ \bibnamefont {Bodmer}},\ }\bibfield  {title} {\bibinfo {title} {{Relativistic Calculation of Nuclear Matter and the Nuclear Surface}},\ }\href {https://doi.org/10.1016/0375-9474(77)90626-1} {\bibfield  {journal} {\bibinfo  {journal} {Nucl. Phys. A}\ }\textbf {\bibinfo {volume} {292}},\ \bibinfo {pages} {413} (\bibinfo {year} {1977})}\BibitemShut {NoStop}%
\bibitem [{\citenamefont {Motohiro}\ \emph {et~al.}(2015)\citenamefont {Motohiro}, \citenamefont {Kim},\ and\ \citenamefont {Harada}}]{Motohiro:2015taa}%
  \BibitemOpen
  \bibfield  {author} {\bibinfo {author} {\bibfnamefont {Y.}~\bibnamefont {Motohiro}}, \bibinfo {author} {\bibfnamefont {Y.}~\bibnamefont {Kim}},\ and\ \bibinfo {author} {\bibfnamefont {M.}~\bibnamefont {Harada}},\ }\bibfield  {title} {\bibinfo {title} {{Asymmetric nuclear matter in a parity doublet model with hidden local symmetry}},\ }\href {https://doi.org/10.1103/PhysRevC.92.025201} {\bibfield  {journal} {\bibinfo  {journal} {Phys. Rev. C}\ }\textbf {\bibinfo {volume} {92}},\ \bibinfo {pages} {025201} (\bibinfo {year} {2015})},\ \bibinfo {note} {[Erratum: Phys.Rev.C 95, 059903 (2017)]},\ \Eprint {https://arxiv.org/abs/1505.00988} {arXiv:1505.00988 [nucl-th]} \BibitemShut {NoStop}%
\bibitem [{\citenamefont {Malik}\ \emph {et~al.}(2024)\citenamefont {Malik}, \citenamefont {Dexheimer},\ and\ \citenamefont {Provid{\^e}ncia}}]{Malik:2024qjw}%
  \BibitemOpen
  \bibfield  {author} {\bibinfo {author} {\bibfnamefont {T.}~\bibnamefont {Malik}}, \bibinfo {author} {\bibfnamefont {V.}~\bibnamefont {Dexheimer}},\ and\ \bibinfo {author} {\bibfnamefont {C.}~\bibnamefont {Provid{\^e}ncia}},\ }\bibfield  {title} {\bibinfo {title} {{Astrophysics and nuclear physics informed interactions in dense matter: Inclusion of PSR J0437-4715}},\ }\href {https://doi.org/10.1103/PhysRevD.110.043042} {\bibfield  {journal} {\bibinfo  {journal} {Phys. Rev. D}\ }\textbf {\bibinfo {volume} {110}},\ \bibinfo {pages} {043042} (\bibinfo {year} {2024})},\ \Eprint {https://arxiv.org/abs/2404.07936} {arXiv:2404.07936 [nucl-th]} \BibitemShut {NoStop}%
\bibitem [{\citenamefont {Drischler}\ \emph {et~al.}(2020)\citenamefont {Drischler}, \citenamefont {Melendez}, \citenamefont {Furnstahl},\ and\ \citenamefont {Phillips}}]{Drischler:2020yad}%
  \BibitemOpen
  \bibfield  {author} {\bibinfo {author} {\bibfnamefont {C.}~\bibnamefont {Drischler}}, \bibinfo {author} {\bibfnamefont {J.~A.}\ \bibnamefont {Melendez}}, \bibinfo {author} {\bibfnamefont {R.~J.}\ \bibnamefont {Furnstahl}},\ and\ \bibinfo {author} {\bibfnamefont {D.~R.}\ \bibnamefont {Phillips}},\ }\bibfield  {title} {\bibinfo {title} {{Quantifying uncertainties and correlations in the nuclear-matter equation of state}},\ }\href {https://doi.org/10.1103/PhysRevC.102.054315} {\bibfield  {journal} {\bibinfo  {journal} {Phys. Rev. C}\ }\textbf {\bibinfo {volume} {102}},\ \bibinfo {pages} {054315} (\bibinfo {year} {2020})},\ \Eprint {https://arxiv.org/abs/2004.07805} {arXiv:2004.07805 [nucl-th]} \BibitemShut {NoStop}%
\bibitem [{\citenamefont {Miller}\ \emph {et~al.}(2019)\citenamefont {Miller} \emph {et~al.}}]{Miller:2019cac}%
  \BibitemOpen
  \bibfield  {author} {\bibinfo {author} {\bibfnamefont {M.~C.}\ \bibnamefont {Miller}} \emph {et~al.},\ }\bibfield  {title} {\bibinfo {title} {{PSR J0030+0451 Mass and Radius from $NICER$ Data and Implications for the Properties of Neutron Star Matter}},\ }\href {https://doi.org/10.3847/2041-8213/ab50c5} {\bibfield  {journal} {\bibinfo  {journal} {Astrophys. J. Lett.}\ }\textbf {\bibinfo {volume} {887}},\ \bibinfo {pages} {L24} (\bibinfo {year} {2019})},\ \Eprint {https://arxiv.org/abs/1912.05705} {arXiv:1912.05705 [astro-ph.HE]} \BibitemShut {NoStop}%
\bibitem [{\citenamefont {Riley}\ \emph {et~al.}(2019)\citenamefont {Riley} \emph {et~al.}}]{Riley:2019yda}%
  \BibitemOpen
  \bibfield  {author} {\bibinfo {author} {\bibfnamefont {T.~E.}\ \bibnamefont {Riley}} \emph {et~al.},\ }\bibfield  {title} {\bibinfo {title} {{A $NICER$ View of PSR J0030+0451: Millisecond Pulsar Parameter Estimation}},\ }\href {https://doi.org/10.3847/2041-8213/ab481c} {\bibfield  {journal} {\bibinfo  {journal} {Astrophys. J. Lett.}\ }\textbf {\bibinfo {volume} {887}},\ \bibinfo {pages} {L21} (\bibinfo {year} {2019})},\ \Eprint {https://arxiv.org/abs/1912.05702} {arXiv:1912.05702 [astro-ph.HE]} \BibitemShut {NoStop}%
\bibitem [{\citenamefont {Miller}\ \emph {et~al.}(2021)\citenamefont {Miller} \emph {et~al.}}]{Miller:2021qha}%
  \BibitemOpen
  \bibfield  {author} {\bibinfo {author} {\bibfnamefont {M.~C.}\ \bibnamefont {Miller}} \emph {et~al.},\ }\bibfield  {title} {\bibinfo {title} {{The Radius of PSR J0740+6620 from NICER and XMM-Newton Data}},\ }\href {https://doi.org/10.3847/2041-8213/ac089b} {\bibfield  {journal} {\bibinfo  {journal} {Astrophys. J. Lett.}\ }\textbf {\bibinfo {volume} {918}},\ \bibinfo {pages} {L28} (\bibinfo {year} {2021})},\ \Eprint {https://arxiv.org/abs/2105.06979} {arXiv:2105.06979 [astro-ph.HE]} \BibitemShut {NoStop}%
\bibitem [{\citenamefont {Riley}\ \emph {et~al.}(2021)\citenamefont {Riley} \emph {et~al.}}]{Riley:2021pdl}%
  \BibitemOpen
  \bibfield  {author} {\bibinfo {author} {\bibfnamefont {T.~E.}\ \bibnamefont {Riley}} \emph {et~al.},\ }\bibfield  {title} {\bibinfo {title} {{A NICER View of the Massive Pulsar PSR J0740+6620 Informed by Radio Timing and XMM-Newton Spectroscopy}},\ }\href {https://doi.org/10.3847/2041-8213/ac0a81} {\bibfield  {journal} {\bibinfo  {journal} {Astrophys. J. Lett.}\ }\textbf {\bibinfo {volume} {918}},\ \bibinfo {pages} {L27} (\bibinfo {year} {2021})},\ \Eprint {https://arxiv.org/abs/2105.06980} {arXiv:2105.06980 [astro-ph.HE]} \BibitemShut {NoStop}%
\bibitem [{\citenamefont {Choudhury}\ \emph {et~al.}(2024)\citenamefont {Choudhury} \emph {et~al.}}]{Choudhury:2024xbk}%
  \BibitemOpen
  \bibfield  {author} {\bibinfo {author} {\bibfnamefont {D.}~\bibnamefont {Choudhury}} \emph {et~al.},\ }\bibfield  {title} {\bibinfo {title} {{A NICER View of the Nearest and Brightest Millisecond Pulsar: PSR J0437\textendash{}4715}},\ }\href {https://doi.org/10.3847/2041-8213/ad5a6f} {\bibfield  {journal} {\bibinfo  {journal} {Astrophys. J. Lett.}\ }\textbf {\bibinfo {volume} {971}},\ \bibinfo {pages} {L20} (\bibinfo {year} {2024})},\ \Eprint {https://arxiv.org/abs/2407.06789} {arXiv:2407.06789 [astro-ph.HE]} \BibitemShut {NoStop}%
\bibitem [{\citenamefont {Salmi}\ \emph {et~al.}(2022)\citenamefont {Salmi} \emph {et~al.}}]{Salmi:2022cgy}%
  \BibitemOpen
  \bibfield  {author} {\bibinfo {author} {\bibfnamefont {T.}~\bibnamefont {Salmi}} \emph {et~al.},\ }\bibfield  {title} {\bibinfo {title} {{The Radius of PSR J0740+6620 from NICER with NICER Background Estimates}},\ }\href {https://doi.org/10.3847/1538-4357/ac983d} {\bibfield  {journal} {\bibinfo  {journal} {Astrophys. J.}\ }\textbf {\bibinfo {volume} {941}},\ \bibinfo {pages} {150} (\bibinfo {year} {2022})},\ \Eprint {https://arxiv.org/abs/2209.12840} {arXiv:2209.12840 [astro-ph.HE]} \BibitemShut {NoStop}%
\bibitem [{\citenamefont {Salmi}\ \emph {et~al.}(2024)\citenamefont {Salmi} \emph {et~al.}}]{Salmi:2024aum}%
  \BibitemOpen
  \bibfield  {author} {\bibinfo {author} {\bibfnamefont {T.}~\bibnamefont {Salmi}} \emph {et~al.},\ }\bibfield  {title} {\bibinfo {title} {{The Radius of the High-mass Pulsar PSR J0740+6620 with 3.6 yr of NICER Data}},\ }\href {https://doi.org/10.3847/1538-4357/ad5f1f} {\bibfield  {journal} {\bibinfo  {journal} {Astrophys. J.}\ }\textbf {\bibinfo {volume} {974}},\ \bibinfo {pages} {294} (\bibinfo {year} {2024})},\ \Eprint {https://arxiv.org/abs/2406.14466} {arXiv:2406.14466 [astro-ph.HE]} \BibitemShut {NoStop}%
\bibitem [{\citenamefont {Dittmann}\ \emph {et~al.}(2024)\citenamefont {Dittmann} \emph {et~al.}}]{Dittmann:2024mbo}%
  \BibitemOpen
  \bibfield  {author} {\bibinfo {author} {\bibfnamefont {A.~J.}\ \bibnamefont {Dittmann}} \emph {et~al.},\ }\bibfield  {title} {\bibinfo {title} {{A More Precise Measurement of the Radius of PSR J0740+6620 Using Updated NICER Data}},\ }\href {https://doi.org/10.3847/1538-4357/ad5f1e} {\bibfield  {journal} {\bibinfo  {journal} {Astrophys. J.}\ }\textbf {\bibinfo {volume} {974}},\ \bibinfo {pages} {295} (\bibinfo {year} {2024})},\ \Eprint {https://arxiv.org/abs/2406.14467} {arXiv:2406.14467 [astro-ph.HE]} \BibitemShut {NoStop}%
\bibitem [{\citenamefont {Fonseca}\ \emph {et~al.}(2021)\citenamefont {Fonseca} \emph {et~al.}}]{Fonseca:2021wxt}%
  \BibitemOpen
  \bibfield  {author} {\bibinfo {author} {\bibfnamefont {E.}~\bibnamefont {Fonseca}} \emph {et~al.},\ }\bibfield  {title} {\bibinfo {title} {{Refined Mass and Geometric Measurements of the High-mass PSR J0740+6620}},\ }\href {https://doi.org/10.3847/2041-8213/ac03b8} {\bibfield  {journal} {\bibinfo  {journal} {Astrophys. J. Lett.}\ }\textbf {\bibinfo {volume} {915}},\ \bibinfo {pages} {L12} (\bibinfo {year} {2021})},\ \Eprint {https://arxiv.org/abs/2104.00880} {arXiv:2104.00880 [astro-ph.HE]} \BibitemShut {NoStop}%
\bibitem [{\citenamefont {Vinciguerra}\ \emph {et~al.}(2024)\citenamefont {Vinciguerra} \emph {et~al.}}]{Vinciguerra:2023qxq}%
  \BibitemOpen
  \bibfield  {author} {\bibinfo {author} {\bibfnamefont {S.}~\bibnamefont {Vinciguerra}} \emph {et~al.},\ }\bibfield  {title} {\bibinfo {title} {{An Updated Mass\textendash{}Radius Analysis of the 2017\textendash{}2018 NICER Data Set of PSR J0030+0451}},\ }\href {https://doi.org/10.3847/1538-4357/acfb83} {\bibfield  {journal} {\bibinfo  {journal} {Astrophys. J.}\ }\textbf {\bibinfo {volume} {961}},\ \bibinfo {pages} {62} (\bibinfo {year} {2024})},\ \Eprint {https://arxiv.org/abs/2308.09469} {arXiv:2308.09469 [astro-ph.HE]} \BibitemShut {NoStop}%
\bibitem [{\citenamefont {Schmitt}(2010)}]{Schmitt:2010pn}%
  \BibitemOpen
  \bibfield  {author} {\bibinfo {author} {\bibfnamefont {A.}~\bibnamefont {Schmitt}},\ }\href {https://doi.org/10.1007/978-3-642-12866-0} {\emph {\bibinfo {title} {{Dense matter in compact stars: A pedagogical introduction}}}},\ Vol.\ \bibinfo {volume} {811}\ (\bibinfo  {publisher} {Springer Berlin, Heidelberg},\ \bibinfo {year} {2010})\ \Eprint {https://arxiv.org/abs/1001.3294} {arXiv:1001.3294 [astro-ph.SR]} \BibitemShut {NoStop}%
\bibitem [{\citenamefont {Yakovlev}\ \emph {et~al.}(2001)\citenamefont {Yakovlev}, \citenamefont {Kaminker}, \citenamefont {Gnedin},\ and\ \citenamefont {Haensel}}]{Yakovlev:2000jp}%
  \BibitemOpen
  \bibfield  {author} {\bibinfo {author} {\bibfnamefont {D.~G.}\ \bibnamefont {Yakovlev}}, \bibinfo {author} {\bibfnamefont {A.~D.}\ \bibnamefont {Kaminker}}, \bibinfo {author} {\bibfnamefont {O.~Y.}\ \bibnamefont {Gnedin}},\ and\ \bibinfo {author} {\bibfnamefont {P.}~\bibnamefont {Haensel}},\ }\bibfield  {title} {\bibinfo {title} {{Neutrino emission from neutron stars}},\ }\href {https://doi.org/10.1016/S0370-1573(00)00131-9} {\bibfield  {journal} {\bibinfo  {journal} {Phys. Rept.}\ }\textbf {\bibinfo {volume} {354}},\ \bibinfo {pages} {1} (\bibinfo {year} {2001})},\ \Eprint {https://arxiv.org/abs/astro-ph/0012122} {arXiv:astro-ph/0012122} \BibitemShut {NoStop}%
\bibitem [{\citenamefont {Lattimer}\ \emph {et~al.}(1991)\citenamefont {Lattimer}, \citenamefont {Prakash}, \citenamefont {Pethick},\ and\ \citenamefont {Haensel}}]{Lattimer:1991ib}%
  \BibitemOpen
  \bibfield  {author} {\bibinfo {author} {\bibfnamefont {J.~M.}\ \bibnamefont {Lattimer}}, \bibinfo {author} {\bibfnamefont {M.}~\bibnamefont {Prakash}}, \bibinfo {author} {\bibfnamefont {C.~J.}\ \bibnamefont {Pethick}},\ and\ \bibinfo {author} {\bibfnamefont {P.}~\bibnamefont {Haensel}},\ }\bibfield  {title} {\bibinfo {title} {{Direct URCA process in neutron stars}},\ }\href {https://doi.org/10.1103/PhysRevLett.66.2701} {\bibfield  {journal} {\bibinfo  {journal} {Phys. Rev. Lett.}\ }\textbf {\bibinfo {volume} {66}},\ \bibinfo {pages} {2701} (\bibinfo {year} {1991})}\BibitemShut {NoStop}%
\bibitem [{\citenamefont {Friman}\ and\ \citenamefont {Maxwell}(1979)}]{Friman:1979ecl}%
  \BibitemOpen
  \bibfield  {author} {\bibinfo {author} {\bibfnamefont {B.~L.}\ \bibnamefont {Friman}}\ and\ \bibinfo {author} {\bibfnamefont {O.~V.}\ \bibnamefont {Maxwell}},\ }\bibfield  {title} {\bibinfo {title} {{Neutron Star Neutrino Emissivities}},\ }\href {https://doi.org/10.1086/157313} {\bibfield  {journal} {\bibinfo  {journal} {Astrophys. J.}\ }\textbf {\bibinfo {volume} {232}},\ \bibinfo {pages} {541} (\bibinfo {year} {1979})}\BibitemShut {NoStop}%
\bibitem [{\citenamefont {Yakovlev}\ \emph {et~al.}(2004)\citenamefont {Yakovlev}, \citenamefont {Gnedin}, \citenamefont {Kaminker}, \citenamefont {Levenfish},\ and\ \citenamefont {Potekhin}}]{Yakovlev:2003qy}%
  \BibitemOpen
  \bibfield  {author} {\bibinfo {author} {\bibfnamefont {D.~G.}\ \bibnamefont {Yakovlev}}, \bibinfo {author} {\bibfnamefont {O.~Y.}\ \bibnamefont {Gnedin}}, \bibinfo {author} {\bibfnamefont {A.~D.}\ \bibnamefont {Kaminker}}, \bibinfo {author} {\bibfnamefont {K.~P.}\ \bibnamefont {Levenfish}},\ and\ \bibinfo {author} {\bibfnamefont {A.~Y.}\ \bibnamefont {Potekhin}},\ }\bibfield  {title} {\bibinfo {title} {{Neutron star cooling: Theoretical aspects and observational constraints}},\ }\href {https://doi.org/10.1016/j.asr.2003.07.020} {\bibfield  {journal} {\bibinfo  {journal} {Adv. Space Res.}\ }\textbf {\bibinfo {volume} {33}},\ \bibinfo {pages} {523} (\bibinfo {year} {2004})},\ \Eprint {https://arxiv.org/abs/astro-ph/0306143} {arXiv:astro-ph/0306143} \BibitemShut {NoStop}%
\bibitem [{\citenamefont {Grill}\ \emph {et~al.}(2014)\citenamefont {Grill}, \citenamefont {Pais}, \citenamefont {Provid{\^e}ncia}, \citenamefont {Vida{\~na}},\ and\ \citenamefont {Avancini}}]{Grill:2014aea}%
  \BibitemOpen
  \bibfield  {author} {\bibinfo {author} {\bibfnamefont {F.}~\bibnamefont {Grill}}, \bibinfo {author} {\bibfnamefont {H.}~\bibnamefont {Pais}}, \bibinfo {author} {\bibfnamefont {C.}~\bibnamefont {Provid{\^e}ncia}}, \bibinfo {author} {\bibfnamefont {I.}~\bibnamefont {Vida{\~na}}},\ and\ \bibinfo {author} {\bibfnamefont {S.~S.}\ \bibnamefont {Avancini}},\ }\bibfield  {title} {\bibinfo {title} {{Equation of state and thickness of the inner crust of neutron stars}},\ }\href {https://doi.org/10.1103/PhysRevC.90.045803} {\bibfield  {journal} {\bibinfo  {journal} {Phys. Rev. C}\ }\textbf {\bibinfo {volume} {90}},\ \bibinfo {pages} {045803} (\bibinfo {year} {2014})},\ \Eprint {https://arxiv.org/abs/1404.2753} {arXiv:1404.2753 [nucl-th]} \BibitemShut {NoStop}%
\bibitem [{tm1()}]{tm1e_crust}%
  \BibitemOpen
  \href@noop {} {}\bibinfo {howpublished} {\url{https://compose.obspm.fr/eos/207}}\BibitemShut {NoStop}%
\bibitem [{\citenamefont {Alford}\ \emph {et~al.}(2022)\citenamefont {Alford}, \citenamefont {Brodie}, \citenamefont {Haber},\ and\ \citenamefont {Tews}}]{Alford:2022bpp}%
  \BibitemOpen
  \bibfield  {author} {\bibinfo {author} {\bibfnamefont {M.~G.}\ \bibnamefont {Alford}}, \bibinfo {author} {\bibfnamefont {L.}~\bibnamefont {Brodie}}, \bibinfo {author} {\bibfnamefont {A.}~\bibnamefont {Haber}},\ and\ \bibinfo {author} {\bibfnamefont {I.}~\bibnamefont {Tews}},\ }\bibfield  {title} {\bibinfo {title} {{Relativistic mean-field theories for neutron-star physics based on chiral effective field theory}},\ }\href {https://doi.org/10.1103/PhysRevC.106.055804} {\bibfield  {journal} {\bibinfo  {journal} {Phys. Rev. C}\ }\textbf {\bibinfo {volume} {106}},\ \bibinfo {pages} {055804} (\bibinfo {year} {2022})},\ \Eprint {https://arxiv.org/abs/2205.10283} {arXiv:2205.10283 [nucl-th]} \BibitemShut {NoStop}%
\bibitem [{\citenamefont {Prakash}\ \emph {et~al.}(1992)\citenamefont {Prakash}, \citenamefont {Prakash}, \citenamefont {Lattimer},\ and\ \citenamefont {Pethick}}]{Prakash:1992zng}%
  \BibitemOpen
  \bibfield  {author} {\bibinfo {author} {\bibfnamefont {M.}~\bibnamefont {Prakash}}, \bibinfo {author} {\bibfnamefont {M.}~\bibnamefont {Prakash}}, \bibinfo {author} {\bibfnamefont {J.~M.}\ \bibnamefont {Lattimer}},\ and\ \bibinfo {author} {\bibfnamefont {C.~J.}\ \bibnamefont {Pethick}},\ }\bibfield  {title} {\bibinfo {title} {{Rapid cooling of neutron stars by hyperons and Delta isobars}},\ }\href {https://doi.org/10.1086/186376} {\bibfield  {journal} {\bibinfo  {journal} {Astrophys. J. Lett.}\ }\textbf {\bibinfo {volume} {390}},\ \bibinfo {pages} {L77} (\bibinfo {year} {1992})}\BibitemShut {NoStop}%
\bibitem [{\citenamefont {Fattoyev}\ \emph {et~al.}(2010)\citenamefont {Fattoyev}, \citenamefont {Horowitz}, \citenamefont {Piekarewicz},\ and\ \citenamefont {Shen}}]{Fattoyev:2010mx}%
  \BibitemOpen
  \bibfield  {author} {\bibinfo {author} {\bibfnamefont {F.~J.}\ \bibnamefont {Fattoyev}}, \bibinfo {author} {\bibfnamefont {C.~J.}\ \bibnamefont {Horowitz}}, \bibinfo {author} {\bibfnamefont {J.}~\bibnamefont {Piekarewicz}},\ and\ \bibinfo {author} {\bibfnamefont {G.}~\bibnamefont {Shen}},\ }\bibfield  {title} {\bibinfo {title} {{Relativistic effective interaction for nuclei, giant resonances, and neutron stars}},\ }\href {https://doi.org/10.1103/PhysRevC.82.055803} {\bibfield  {journal} {\bibinfo  {journal} {Phys. Rev. C}\ }\textbf {\bibinfo {volume} {82}},\ \bibinfo {pages} {055803} (\bibinfo {year} {2010})},\ \Eprint {https://arxiv.org/abs/1008.3030} {arXiv:1008.3030 [nucl-th]} \BibitemShut {NoStop}%
\bibitem [{\citenamefont {Steinheimer}\ \emph {et~al.}(2025)\citenamefont {Steinheimer}, \citenamefont {Omana~Kuttan}, \citenamefont {Reichert}, \citenamefont {Nara},\ and\ \citenamefont {Bleicher}}]{Steinheimer:2025hsr}%
  \BibitemOpen
  \bibfield  {author} {\bibinfo {author} {\bibfnamefont {J.}~\bibnamefont {Steinheimer}}, \bibinfo {author} {\bibfnamefont {M.}~\bibnamefont {Omana~Kuttan}}, \bibinfo {author} {\bibfnamefont {T.}~\bibnamefont {Reichert}}, \bibinfo {author} {\bibfnamefont {Y.}~\bibnamefont {Nara}},\ and\ \bibinfo {author} {\bibfnamefont {M.}~\bibnamefont {Bleicher}},\ }\bibfield  {title} {\bibinfo {title} {{Simultaneous description of high density QCD matter in heavy ion collisions and neutron star observations}},\ }\href {https://doi.org/10.1016/j.physletb.2025.139605} {\bibfield  {journal} {\bibinfo  {journal} {Phys. Lett. B}\ }\textbf {\bibinfo {volume} {867}},\ \bibinfo {pages} {139605} (\bibinfo {year} {2025})},\ \Eprint {https://arxiv.org/abs/2501.12849} {arXiv:2501.12849 [hep-ph]} \BibitemShut {NoStop}%
\bibitem [{\citenamefont {Brodie}\ \emph {et~al.}(2025)\citenamefont {Brodie}, \citenamefont {Dexheimer}, \citenamefont {Negreiros}, \citenamefont {Pisarski},\ and\ \citenamefont {Steinheimer}}]{PDMCoolingProj}%
  \BibitemOpen
  \bibfield  {author} {\bibinfo {author} {\bibfnamefont {L.}~\bibnamefont {Brodie}}, \bibinfo {author} {\bibfnamefont {V.}~\bibnamefont {Dexheimer}}, \bibinfo {author} {\bibfnamefont {R.}~\bibnamefont {Negreiros}}, \bibinfo {author} {\bibfnamefont {R.~D.}\ \bibnamefont {Pisarski}},\ and\ \bibinfo {author} {\bibfnamefont {J.}~\bibnamefont {Steinheimer}},\ }\href@noop {} {} (\bibinfo {year} {2025}),\ \bibinfo {note} {in progress}\BibitemShut {NoStop}%
\bibitem [{\citenamefont {Brodie}\ and\ \citenamefont {Pisarski}(2025)}]{BrodiePisarski2025Data}%
  \BibitemOpen
  \bibfield  {author} {\bibinfo {author} {\bibfnamefont {L.}~\bibnamefont {Brodie}}\ and\ \bibinfo {author} {\bibfnamefont {R.~D.}\ \bibnamefont {Pisarski}},\ }\href@noop {} {}\bibinfo {howpublished} {\url{https://gitlab.com/liambrodie/neutrino-cooling-from-parity-doubled-nucleons}} (\bibinfo {year} {2025}),\ \bibinfo {note} {{APS} Data Availability}\BibitemShut {NoStop}%
\bibitem [{\citenamefont {Alford}\ \emph {et~al.}(2018)\citenamefont {Alford}, \citenamefont {Bovard}, \citenamefont {Hanauske}, \citenamefont {Rezzolla},\ and\ \citenamefont {Schwenzer}}]{Alford:2017rxf}%
  \BibitemOpen
  \bibfield  {author} {\bibinfo {author} {\bibfnamefont {M.~G.}\ \bibnamefont {Alford}}, \bibinfo {author} {\bibfnamefont {L.}~\bibnamefont {Bovard}}, \bibinfo {author} {\bibfnamefont {M.}~\bibnamefont {Hanauske}}, \bibinfo {author} {\bibfnamefont {L.}~\bibnamefont {Rezzolla}},\ and\ \bibinfo {author} {\bibfnamefont {K.}~\bibnamefont {Schwenzer}},\ }\bibfield  {title} {\bibinfo {title} {{Viscous Dissipation and Heat Conduction in Binary Neutron-Star Mergers}},\ }\href {https://doi.org/10.1103/PhysRevLett.120.041101} {\bibfield  {journal} {\bibinfo  {journal} {Phys. Rev. Lett.}\ }\textbf {\bibinfo {volume} {120}},\ \bibinfo {pages} {041101} (\bibinfo {year} {2018})},\ \Eprint {https://arxiv.org/abs/1707.09475} {arXiv:1707.09475 [gr-qc]} \BibitemShut {NoStop}%
\bibitem [{\citenamefont {Most}\ \emph {et~al.}(2024)\citenamefont {Most}, \citenamefont {Haber}, \citenamefont {Harris}, \citenamefont {Zhang}, \citenamefont {Alford},\ and\ \citenamefont {Noronha}}]{Most:2022yhe}%
  \BibitemOpen
  \bibfield  {author} {\bibinfo {author} {\bibfnamefont {E.~R.}\ \bibnamefont {Most}}, \bibinfo {author} {\bibfnamefont {A.}~\bibnamefont {Haber}}, \bibinfo {author} {\bibfnamefont {S.~P.}\ \bibnamefont {Harris}}, \bibinfo {author} {\bibfnamefont {Z.}~\bibnamefont {Zhang}}, \bibinfo {author} {\bibfnamefont {M.~G.}\ \bibnamefont {Alford}},\ and\ \bibinfo {author} {\bibfnamefont {J.}~\bibnamefont {Noronha}},\ }\bibfield  {title} {\bibinfo {title} {{Emergence of Microphysical Bulk Viscosity in Binary Neutron Star Postmerger Dynamics}},\ }\href {https://doi.org/10.3847/2041-8213/ad454f} {\bibfield  {journal} {\bibinfo  {journal} {Astrophys. J. Lett.}\ }\textbf {\bibinfo {volume} {967}},\ \bibinfo {pages} {L14} (\bibinfo {year} {2024})},\ \Eprint {https://arxiv.org/abs/2207.00442} {arXiv:2207.00442 [astro-ph.HE]} \BibitemShut {NoStop}%
\bibitem [{\citenamefont {Heinke}\ \emph {et~al.}(2007)\citenamefont {Heinke}, \citenamefont {Jonker}, \citenamefont {Wijnands},\ and\ \citenamefont {Taam}}]{Heinke:2006ie}%
  \BibitemOpen
  \bibfield  {author} {\bibinfo {author} {\bibfnamefont {C.~O.}\ \bibnamefont {Heinke}}, \bibinfo {author} {\bibfnamefont {P.~G.}\ \bibnamefont {Jonker}}, \bibinfo {author} {\bibfnamefont {R.}~\bibnamefont {Wijnands}},\ and\ \bibinfo {author} {\bibfnamefont {R.~E.}\ \bibnamefont {Taam}},\ }\bibfield  {title} {\bibinfo {title} {{Constraints on Thermal X-ray Radiation from SAX J1808.4-3658 and Implications for Neutron Star Neutrino Emission}},\ }\href {https://doi.org/10.1086/513140} {\bibfield  {journal} {\bibinfo  {journal} {Astrophys. J.}\ }\textbf {\bibinfo {volume} {660}},\ \bibinfo {pages} {1424} (\bibinfo {year} {2007})},\ \Eprint {https://arxiv.org/abs/astro-ph/0612232} {arXiv:astro-ph/0612232} \BibitemShut {NoStop}%
\bibitem [{\citenamefont {Jonker}\ \emph {et~al.}(2006)\citenamefont {Jonker}, \citenamefont {Bassa}, \citenamefont {Nelemans}, \citenamefont {Juett}, \citenamefont {Brown},\ and\ \citenamefont {Chakrabarty}}]{Jonker:2006td}%
  \BibitemOpen
  \bibfield  {author} {\bibinfo {author} {\bibfnamefont {P.~G.}\ \bibnamefont {Jonker}}, \bibinfo {author} {\bibfnamefont {C.~G.}\ \bibnamefont {Bassa}}, \bibinfo {author} {\bibfnamefont {G.}~\bibnamefont {Nelemans}}, \bibinfo {author} {\bibfnamefont {A.~M.}\ \bibnamefont {Juett}}, \bibinfo {author} {\bibfnamefont {E.~F.}\ \bibnamefont {Brown}},\ and\ \bibinfo {author} {\bibfnamefont {D.}~\bibnamefont {Chakrabarty}},\ }\bibfield  {title} {\bibinfo {title} {{The neutron star soft x-ray transient 1h1905+000 in quiescence}},\ }\href {https://doi.org/10.1111/j.1365-2966.2006.10253.x} {\bibfield  {journal} {\bibinfo  {journal} {Mon. Not. Roy. Astron. Soc.}\ }\textbf {\bibinfo {volume} {368}},\ \bibinfo {pages} {1803} (\bibinfo {year} {2006})},\ \Eprint {https://arxiv.org/abs/astro-ph/0602625} {arXiv:astro-ph/0602625} \BibitemShut {NoStop}%
\bibitem [{\citenamefont {Navas}\ \emph {et~al.}(2024)\citenamefont {Navas} \emph {et~al.}}]{ParticleDataGroup:2024cfk}%
  \BibitemOpen
  \bibfield  {author} {\bibinfo {author} {\bibfnamefont {S.}~\bibnamefont {Navas}} \emph {et~al.} (\bibinfo {collaboration} {Particle Data Group}),\ }\bibfield  {title} {\bibinfo {title} {{Review of particle physics}},\ }\href {https://doi.org/10.1103/PhysRevD.110.030001} {\bibfield  {journal} {\bibinfo  {journal} {Phys. Rev. D}\ }\textbf {\bibinfo {volume} {110}},\ \bibinfo {pages} {030001} (\bibinfo {year} {2024})}\BibitemShut {NoStop}%
\bibitem [{\citenamefont {Page}\ \emph {et~al.}(2009)\citenamefont {Page}, \citenamefont {Lattimer}, \citenamefont {Prakash},\ and\ \citenamefont {Steiner}}]{Page:2009fu}%
  \BibitemOpen
  \bibfield  {author} {\bibinfo {author} {\bibfnamefont {D.}~\bibnamefont {Page}}, \bibinfo {author} {\bibfnamefont {J.~M.}\ \bibnamefont {Lattimer}}, \bibinfo {author} {\bibfnamefont {M.}~\bibnamefont {Prakash}},\ and\ \bibinfo {author} {\bibfnamefont {A.~W.}\ \bibnamefont {Steiner}},\ }\bibfield  {title} {\bibinfo {title} {{Neutrino Emission from Cooper Pairs and Minimal Cooling of Neutron Stars}},\ }\href {https://doi.org/10.1088/0004-637X/707/2/1131} {\bibfield  {journal} {\bibinfo  {journal} {Astrophys. J.}\ }\textbf {\bibinfo {volume} {707}},\ \bibinfo {pages} {1131} (\bibinfo {year} {2009})},\ \Eprint {https://arxiv.org/abs/0906.1621} {arXiv:0906.1621 [astro-ph.SR]} \BibitemShut {NoStop}%
\bibitem [{\citenamefont {Beloin}\ \emph {et~al.}(2018)\citenamefont {Beloin}, \citenamefont {Han}, \citenamefont {Steiner},\ and\ \citenamefont {Page}}]{Beloin:2016zop}%
  \BibitemOpen
  \bibfield  {author} {\bibinfo {author} {\bibfnamefont {S.}~\bibnamefont {Beloin}}, \bibinfo {author} {\bibfnamefont {S.}~\bibnamefont {Han}}, \bibinfo {author} {\bibfnamefont {A.~W.}\ \bibnamefont {Steiner}},\ and\ \bibinfo {author} {\bibfnamefont {D.}~\bibnamefont {Page}},\ }\bibfield  {title} {\bibinfo {title} {{Constraining Superfluidity in Dense Matter from the Cooling of Isolated Neutron Stars}},\ }\href {https://doi.org/10.1103/PhysRevC.97.015804} {\bibfield  {journal} {\bibinfo  {journal} {Phys. Rev. C}\ }\textbf {\bibinfo {volume} {97}},\ \bibinfo {pages} {015804} (\bibinfo {year} {2018})},\ \Eprint {https://arxiv.org/abs/1612.04289} {arXiv:1612.04289 [nucl-th]} \BibitemShut {NoStop}%
\bibitem [{\citenamefont {Sedrakian}\ and\ \citenamefont {Clark}(2019)}]{Sedrakian:2018ydt}%
  \BibitemOpen
  \bibfield  {author} {\bibinfo {author} {\bibfnamefont {A.}~\bibnamefont {Sedrakian}}\ and\ \bibinfo {author} {\bibfnamefont {J.~W.}\ \bibnamefont {Clark}},\ }\bibfield  {title} {\bibinfo {title} {{Superfluidity in nuclear systems and neutron stars}},\ }\href {https://doi.org/10.1140/epja/i2019-12863-6} {\bibfield  {journal} {\bibinfo  {journal} {Eur. Phys. J. A}\ }\textbf {\bibinfo {volume} {55}},\ \bibinfo {pages} {167} (\bibinfo {year} {2019})},\ \Eprint {https://arxiv.org/abs/1802.00017} {arXiv:1802.00017 [nucl-th]} \BibitemShut {NoStop}%
\bibitem [{\citenamefont {Page}\ \emph {et~al.}(2013)\citenamefont {Page}, \citenamefont {Lattimer}, \citenamefont {Prakash},\ and\ \citenamefont {Steiner}}]{Page:2013hxa}%
  \BibitemOpen
  \bibfield  {author} {\bibinfo {author} {\bibfnamefont {D.}~\bibnamefont {Page}}, \bibinfo {author} {\bibfnamefont {J.~M.}\ \bibnamefont {Lattimer}}, \bibinfo {author} {\bibfnamefont {M.}~\bibnamefont {Prakash}},\ and\ \bibinfo {author} {\bibfnamefont {A.~W.}\ \bibnamefont {Steiner}},\ }\href@noop {} {\bibinfo {title} {Stellar superfluids}} (\bibinfo {year} {2013}),\ \bibinfo {note} {arXiv:1302.6626 [astro-ph.HE], Report: INT-PUB-13-009}\BibitemShut {NoStop}%
\bibitem [{\citenamefont {Brown}\ \emph {et~al.}(2018)\citenamefont {Brown}, \citenamefont {Cumming}, \citenamefont {Fattoyev}, \citenamefont {Horowitz}, \citenamefont {Page},\ and\ \citenamefont {Reddy}}]{Brown:2017gxd}%
  \BibitemOpen
  \bibfield  {author} {\bibinfo {author} {\bibfnamefont {E.~F.}\ \bibnamefont {Brown}}, \bibinfo {author} {\bibfnamefont {A.}~\bibnamefont {Cumming}}, \bibinfo {author} {\bibfnamefont {F.~J.}\ \bibnamefont {Fattoyev}}, \bibinfo {author} {\bibfnamefont {C.~J.}\ \bibnamefont {Horowitz}}, \bibinfo {author} {\bibfnamefont {D.}~\bibnamefont {Page}},\ and\ \bibinfo {author} {\bibfnamefont {S.}~\bibnamefont {Reddy}},\ }\bibfield  {title} {\bibinfo {title} {{Rapid neutrino cooling in the neutron star MXB 1659-29}},\ }\href {https://doi.org/10.1103/PhysRevLett.120.182701} {\bibfield  {journal} {\bibinfo  {journal} {Phys. Rev. Lett.}\ }\textbf {\bibinfo {volume} {120}},\ \bibinfo {pages} {182701} (\bibinfo {year} {2018})},\ \Eprint {https://arxiv.org/abs/1801.00041} {arXiv:1801.00041 [astro-ph.HE]} \BibitemShut {NoStop}%
\bibitem [{\citenamefont {Marino}\ \emph {et~al.}(2024)\citenamefont {Marino}, \citenamefont {Dehman}, \citenamefont {Kovlakas}, \citenamefont {Rea}, \citenamefont {Pons},\ and\ \citenamefont {Vigan{\`o}}}]{Marino:2024gpm}%
  \BibitemOpen
  \bibfield  {author} {\bibinfo {author} {\bibfnamefont {A.}~\bibnamefont {Marino}}, \bibinfo {author} {\bibfnamefont {C.}~\bibnamefont {Dehman}}, \bibinfo {author} {\bibfnamefont {K.}~\bibnamefont {Kovlakas}}, \bibinfo {author} {\bibfnamefont {N.}~\bibnamefont {Rea}}, \bibinfo {author} {\bibfnamefont {J.~A.}\ \bibnamefont {Pons}},\ and\ \bibinfo {author} {\bibfnamefont {D.}~\bibnamefont {Vigan{\`o}}},\ }\bibfield  {title} {\bibinfo {title} {{Constraints on the dense matter equation of state from young and cold isolated neutron stars}},\ }\href {https://doi.org/10.1038/s41550-024-02291-y} {\bibfield  {journal} {\bibinfo  {journal} {Nature Astron.}\ }\textbf {\bibinfo {volume} {8}},\ \bibinfo {pages} {1020} (\bibinfo {year} {2024})},\ \Eprint {https://arxiv.org/abs/2404.05371} {arXiv:2404.05371 [astro-ph.HE]} \BibitemShut {NoStop}%
\end{thebibliography}%
\end{document}